\documentclass[pre, aps, twocolumn, longbibliography, notitlepage, nobibnotes, superscriptaddress,nofootinbib]{revtex4-2}
\usepackage[colorlinks, allcolors=darkblue]{hyperref}
\usepackage{color}
\pdfoutput=1

\usepackage{xr}
\usepackage{graphicx}
\usepackage{dcolumn}
\usepackage{bm}
\usepackage{xcolor}
\definecolor{darkblue}{rgb}{0,0,0.6}
\usepackage[detect-all=true]{siunitx}
\usepackage{physics}
\usepackage{mathtools}
\usepackage{enumitem}

\newcommand{\exgr}[0]{e.g.}
\newcommand{\idest}[0]{i.e.}

\newcommand\articletitle{
Correlated cell movements drive epithelial finger formation}

\newcommand\articleabstract{ 
Sheets of epithelial cells form protective barriers in multicellular organisms. When damaged, finger-like protrusions form at the advancing edge, closing the damaged area. Due to the resemblance to fluid spreading, existing models typically invoke instability mechanisms to explain the onset of fingers. Combining in vitro experiments on freely expanding MDCK cell monolayers with simulations of the self-propelled Voronoi model and an active viscoelastic theory, we show that instead fingers form spontaneously due to emergent, correlated cell motion within the cell layer and simply represent long-lived active fluctuations of the boundary. Simulations and theory both quantitatively match spatiotemporally correlated cell motion measured in the interior of the monolayer. To capture finger formation, we model the actomyosin cable at the advancing edge as a contractile semi-flexible polymer driven by correlated active noise representing the interior. The model not only exhibits spontaneous finger formation but also quantitatively predicts tangent-tangent and roughness correlation functions of the edge in space and time, as well as fluctuation spectra. Our results, therefore, indicate that correlated cell movements lead to robust finger formation, without the need for any feedback mechanism, suggesting that leader cells, cell-cell signalling, and division modulate an intrinsic process instead of causing it.}

\newcommand\articlesignificancestatement{
Epithelia form protective barriers in multicellular organisms. To maintain homeostasis, they must be able to regenerate and heal damaged areas. This occurs through collective cell migration, during which finger-like protrusions commonly appear. Whether these protrusions are driven by specialised leader cells, biochemical cues, or generic physical interactions remains unclear. Integrating in vitro imaging, agent-based simulations, and continuum modelling, we show that correlated active cell motion alone suffices to produce fingers. Leader cells, signalling, and proliferation modulate, but do not trigger, this pattern. Our results show that the key mechanism underlying a complex biological process can be understood using a general framework of the physics of dense active matter.
}

\newcommand\articleauthorcontributions{
SCK carried out the image analysis and ran the simulations. PA and IPN carried out the experiments in the IN lab. TBL and RS co-formulated the polymer model with SCK and SH. RS and SH developed the AVM. SH formulated the research, and developed the theory together with SCK and YEK. SCK, IN, RS, YEK and SH wrote the manuscript.
}

\newcommand\articleacknow{
S.H. would like to acknowledge the NWO Sector plan for providing start-up funding.
This work was performed using the computing resources from the Academic Leiden Interdisciplinary Cluster Environment (ALICE) provided by Leiden University. R.S.\ acknowledges support from the UK Engineering and Physical Sciences Research Council (Award EP/W023946/1). T.B.L.\ was supported by EPSRC grants EP/R014604/1 and EP/T031077/1. S.H. and R.S. would like to acknowledge the 2024 Active Solids KITP program, and this research was supported in part by grant NSF PHY-2309135 to the Kavli Institute for Theoretical Physics (KITP). 
}

\newcommand\dropcap[1]{#1}

\usepackage{tcolorbox}
\newtcolorbox{mybox}{colback=black!5!white,colframe=white!75!black}

\usepackage{ragged2e}

\newcommand{\la}{\left \langle}
\newcommand{\ra}{\right \rangle}
\newcommand{\buv}[1]{\bm{\hat{#1}}}
\newcommand{\bt}[1]{\bm{ \overleftrightarrow{#1} }}
\newcommand{\ii}{\mathrm{i}}

\usepackage{titlesec}

\usepackage{chngcntr}

\begin{document}

\title{\articletitle{}}

\author{Sander C. Kammeraat}
\affiliation{Instituut-Lorentz for Theoretical Physics, Universiteit Leiden, 2333 CA Leiden, Netherlands}

\author{Yann-Edwin Keta}
\affiliation{Instituut-Lorentz for Theoretical Physics, Universiteit Leiden, 2333 CA Leiden, Netherlands}

\author{Paul Appleton}
\affiliation{School of Life Sciences, University of Dundee, Dundee DD1 5EH, United Kingdom}

\author{Ian P. Newton}
\affiliation{School of Life Sciences, University of Dundee, Dundee DD1 5EH, United Kingdom}

\author{Tanniemola B. Liverpool}
\affiliation{School of Mathematics, University of Bristol, Bristol BS8 1UG, UK}

\author{Rastko Sknepnek}
\affiliation{School of Life Sciences, University of Dundee, Dundee DD1 5EH, United Kingdom}
\affiliation{School of Science and Engineering, University of Dundee, Dundee DD1 4HN, United Kingdom}

\author{Inke Näthke}
\affiliation{School of Life Sciences, University of Dundee, Dundee DD1 5EH, United Kingdom}

\author{Silke Henkes}
\affiliation{Instituut-Lorentz for Theoretical Physics, Universiteit Leiden, 2333 CA Leiden, Netherlands}
\email{shenkes@lorentz.leidenuniv.nl}

\begin{abstract}
\articleabstract{}
\footnote{{\bf{Author contributions: }}\articleauthorcontributions{}}
\end{abstract}

\maketitle

\section*{Significance statement}
\articlesignificancestatement{}

\section*{Introduction}

\dropcap{S}tructural integrity of the epithelial layers that form barriers in multicellular organisms \cite{hondaWorldEpithelialSheets2017} is vital for protection against harmful external agents and regulated transport into and out of the body.  Consequently, their ability to heal rapidly when damaged is critical for normal homeostasis, prompting detailed investigations into underlying biological and mechanical mechanisms.   Early studies \cite{nikolicRoleBoundaryConditions2006,poujadeCollectiveMigrationEpithelial2007,petitjeanVelocityFieldsCollectively2010,reffayOrientationPolarityCollectively2011}  using epithelial monolayers revealed that healing does not occur as a flat front moving into the wounded region. Instead, finger-like multicellular structures emerge at the boundary and invade the wounded region. This phenomenon, exhibited in moving epithelial sheets, has been studied extensively \cite{poujadeCollectiveMigrationEpithelial2007,ouakninStochasticCollectiveMovement2009,petitjeanVelocityFieldsCollectively2010,markPhysicalModelDynamic2010,reffayOrientationPolarityCollectively2011,basanAlignmentCellularMotility2013,h.kopfContinuumModelEpithelial2013,sepulvedaCollectiveCellMotion2013,reffayInterplayRhoAMechanical2014,zimmermannInstabilityEdgeTissue2014,tarleModelingFingerInstability2015,vishwakarmaMechanicalInteractionsFollowers2018,alertActiveFingeringInstability2019,yangLeadercelldrivenEpithelialSheet2020,ogumaMechanismUnderlyingDynamic2020,vishwakarmaMechanobiologyLeaderFollower2020,buscherInstabilityFingeringInterfaces2020,trenadoFingeringInstabilitySpreading2021,mukhtarMultiscaleComputationalModel2022,killeenModelingGrowingConfluent2023,jeongFingeringPatternsEpithelial2023,yeFingeringInstabilityAccelerates2024},

producing several different potential mechanisms. Yet, no consensus has been reached about the exact biophysical mechanism underlying this phenomenon.

Several models have been proposed where leader cells at the tips of the fingers, with prominent lamellipodia that generate traction forces \cite{qinRolesLeaderFollower2021, {vishwakarmaMechanobiologyLeaderFollower2020}}, initiate the formation of the fingers \cite{sepulvedaCollectiveCellMotion2013,yangLeadercelldrivenEpithelialSheet2020,ogumaMechanismUnderlyingDynamic2020}.
The interaction between motility and the shape of these finger tips has also been proposed to create an instability through curvature-enhanced motility of cells at the moving boundary~\cite{markPhysicalModelDynamic2010,tarleModelingFingerInstability2015}.
Models introducing signalling \cite{ouakninStochasticCollectiveMovement2009,h.kopfContinuumModelEpithelial2013} between cells and consequent intracellular changes related to, for instance, YAP signalling and other pathways have also been developed \cite{mukhtarMultiscaleComputationalModel2022}, involving complex gene expression changes \cite{qinRolesLeaderFollower2021} in the process.
Incorporating cell proliferation at the advancing edge of the monolayer has led to alternative views on how fingers are formed.  Specifically, these models only focus on the boundary, treating finger formation as an interfacial problem~\cite{rapinRoughnessDynamicsProliferating2021}, either as part of the KPZ universality class~\cite{killeenModelingGrowingConfluent2023} or with interaction between cell growth and pressure \cite{yeFingeringInstabilityAccelerates2024}.

\begin{figure*}
    \centering
    \includegraphics{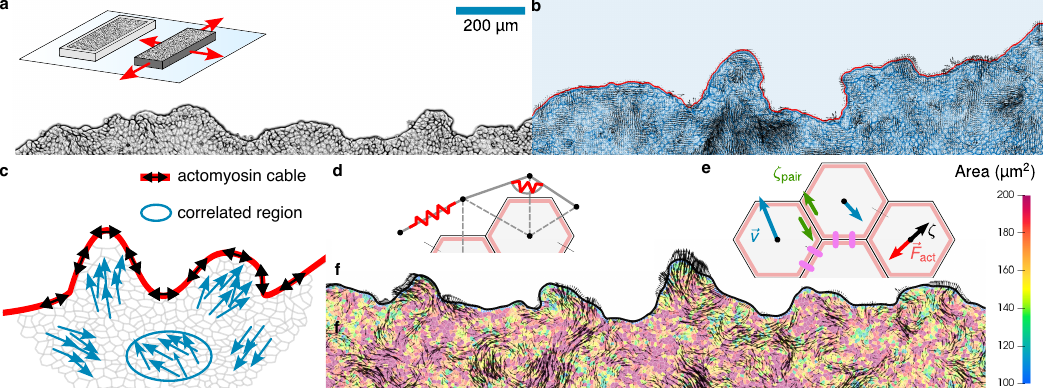}
    \caption{Experiments and model. (\textbf{a}) MDCK cells are cultured in a gasket, which is then removed after the cells have grown to confluence (Materials and Methods). The cells then invade the free space, forming finger-like structures. Scale bar: 200 \si{\micro \meter}. (\textbf{b}) We used a boundary detection algorithm to extract the boundary (red line) and Particle Image Velocimetry (PIV) to obtain the velocity fields (black arrows) from the experiments. (\textbf{c}) We propose a mechanism for the finger formation based on the interplay between correlated cell movements (shown in blue) and the contractile actomyosin cable(s) at the advancing edge (red curve) of the cell monolayer. (\textbf{d}\&\textbf{e}) To test this model, we use an active vertex model that reproduces the velocity correlations and includes a contractile boundary. (\textbf{d}) The contractile boundary consists of coupled linear and angular springs (red springs), with boundary points that can be added or removed. 
    (\textbf{e}) The velocity of the model cells (blue arrows) is set by the balance of dissipative forces, consisting of substrate friction (black arrow, friction coefficient $\zeta$) and dissipation between pairs of cells (green arrows, friction coefficient $\zeta_{\text{pair}}$), with an uncorrelated active crawling force (red arrow) and elastic cell-cell forces, consisting of area and perimeter forces from the standard vertex model potential (Methods), implementing cell-cell adhesion (pink) and cell contractility (salmon).
    (\textbf{f}) Typical simulation snapshot with fingers in steady-state. Model cells are coloured by area, and black arrows indicate cell velocities.}
    \label{fig:overview}
\end{figure*}

Recent studies increasingly focused on the role of the entire monolayer in finger formation, reflecting the increased recognition of the role of mechanics in the collective behaviour of tissues \cite{ladouxMechanobiologyCollectiveCell2017}. 
In particular, it was shown that the mechanism controlling finger formation must originate from the dynamics of cells in the bulk behind the fingers \cite{vishwakarmaMechanicalInteractionsFollowers2018}, and that leader cells cannot drag more than one or two follower cells \cite{rossettiOptogeneticGenerationLeader2024,leeItTakesMore2024}, and that the boundary and interior length scales are coupled \cite{vazquez2022effect}. From a modelling point of view, cell-cell alignment in the interior has been shown to generate finger structures, either on its own~\cite{basanAlignmentCellularMotility2013, khataeeMultiscaleModellingMotility2020, buscherInstabilityFingeringInterfaces2020} or in conjunction with leaders \cite{sepulvedaCollectiveCellMotion2013,tarleModelingFingerInstability2015} or cell division \cite{bonillaTrackingCollectiveCell2020}.
Moreover, at the continuum scale, the interplay between an imposed velocity gradient and contractile intercellular stresses is known to generate boundary instabilities \cite{leeCrawlingCellsCan2011, nesbittEdgeInstabilityIncompressible2017,alertActiveFingeringInstability2019,trenadoFingeringInstabilitySpreading2021}.

In this paper, we propose a simple mechanism for the formation of fingers, as illustrated in Fig.\ \ref{fig:overview}.
Starting from observations of correlated cell movements over length scales exceeding multiple cell sizes \cite{garciaPhysicsActiveJamming2015,petitjeanVelocityFieldsCollectively2010,henkesDenseActiveMatter2020}, we show that uncorrelated active cell crawling, together with elastic interactions between cells, results in multicellular-scale flow patterns in the cell monolayer behind the advancing edge \cite{henkesDenseActiveMatter2020}.  Building on our previous work \cite{henkesDenseActiveMatter2020} and including the mechanical effects of the multicellular actomyosin cable on finger formation \cite{jeongFingeringPatternsEpithelial2023,reffayInterplayRhoAMechanical2014}, we develop a cell-level computational model and a viscoelastic continuum theory of the advancing edge that 
both describe finger formation as the interplay between these correlated cell movements and the contractile actomyosin cable.

Using data from an MDCK wound healing assay imaged over large length scales and long times, we established that fingers slowly develop and reach a steady state, but without the growth of a characteristic length scale typical of a finite-wavelength instability. 
The model and the continuum theory are able to quantitatively reproduce velocity correlations and finger properties observed in experiments. 
Our findings, therefore, indicate that epithelial fingers are highly persistent fluctuations arising from active driving of the advancing edge by the interior of the monolayer.

\section*{Results - monolayer interior}

To understand the origin and behaviour of the fingers, we studied both the properties of the cells in the bulk of a confluent epithelial monolayer and the behaviour of the advancing edge invading the free space. We begin by examining the behaviour in the bulk. 

\subsection*{\label{sec:experiments}Experiments}

\begin{figure*}
    \centering
    \includegraphics[width=1\textwidth]{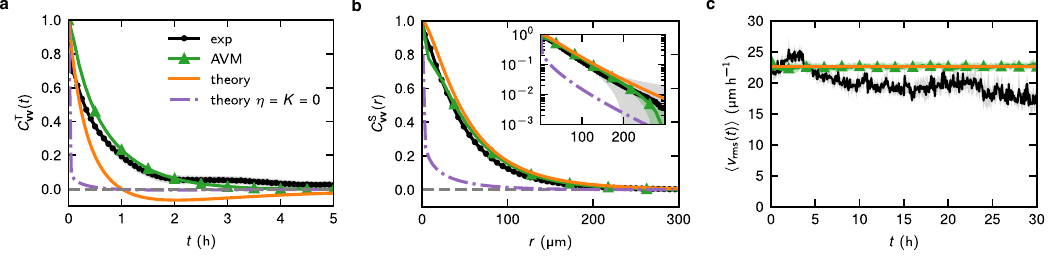}
    \caption{Correlations of the velocity field in the cell sheet for the experiments (black dots), the AVM simulations (green triangles), and the viscoelastic theory either with internal dissipation (orange curve) or without (dashed purple curve). The shading indicates the standard error of the mean. (\textbf{a}) The temporal velocity autocorrelation function measures the persistence of the cell velocities over time. (\textbf{b}) The spatial velocity correlation function measures the spatial extent of the correlated movements. Inset: log-lin scale. (\textbf{c}) Root mean square velocity of the cells. For clarity, we show only some of the markers for the AVM simulations.}
\label{fig:velocity_statistics_panel}
\end{figure*}

We studied the velocity correlations and finger formation experimentally using Madin-Darby canine kidney (MDCK) cell monolayers (Methods). 
After extracting the velocity field using Particle Image Velocimetry (PIV) on these images (SI Appendix, section \ref{subsec:PIV}), and removing the mean velocity vector to correct for stitching artefacts, we calculated various velocity statistics (Fig.\ \ref{fig:velocity_statistics_panel}) to calibrate the Active Vertex Model (AVM) discussed below. We work in a frame that is co-moving with the mean boundary progression. Earlier experiments revealed gradients in the velocity fields as a function of distance to the moving front \cite{petitjeanVelocityFieldsCollectively2010}. To account for this in our analysis, we calculated the velocity statistics at different distances behind the fingers. We confirmed a slight dependency of velocity on location (SI Appendix, section \ref{vel_stat_further_behind}). In the remainder of the paper, we show the velocity statistics for the region immediately behind the fingers (defined in SI Appendix, section \ref{subsec:zoning}), as we used the velocity statistics of this region for the calibration of the AVM parameters.    

We quantified a typical evolution of the boundary contour in an experiment (Fig.\ \ref{fig:contour_panel}a) using the tangent-tangent correlation function (Fig.\ \ref{fig:tangent_spatial_roughness_panel}a), the roughness correlation (Fig.\ \ref{fig:tangent_spatial_roughness_panel}e) and the Fourier spectrum of the boundary (Fig.\ \ref{fig:tangent_spatial_roughness_panel}h).

\subsection*{\label{sec:AVM}Active Vertex Model}
We represent the cell sheet through a version of the vertex model~\cite{nagaiDynamicCellModel2001,farhadifarInfluenceCellMechanics2007,fletcherVertexModelsEpithelial2014}, where the sheet is represented as a two-dimensional Voronoi tiling \cite{liCoherentMotionsConfluent2014,biMotilityDrivenGlassJamming2016,bartonActiveVertexModel2017}, extended to include a dynamic monolayer edge. 
The edge is modelled as a semi-flexible polymer under tension to mimic the contractile nature of the actomyosin cable(s). 
It can dynamically grow and shrink, making it a good computational model to study finger formation. The total energy of our model can be written as the sum of the mechanical energy of cells in the bulk, $E_{\mathrm{VM}}$, with an area term of modulus $K_A$ and a perimeter term of modulus $K_P$, see Eq.\ \eqref{eq:VM}, and the deformation energy of the advancing edge, $E_{\textrm{edge}}$,  with stretching modulus $k_s$ and bending modulus $k_b$, see Eq.\ \eqref{eq:Eedge}, giving in total $E_{\text{AVM}} = E_{\text{VM}} + E_{\text{edge}}$ (see Methods for details).

The model is made active by adding self-propulsion to each cell. Although the individual crawling of epithelial cells in a confluent monolayer on a substrate is a complex process, involving cryptic lamellipodia \cite{farooquiMultipleRowsCells2005}, the observed velocity correlations in the sheet can be successfully captured by assuming that the cells crawl independently as simple Active Brownian Particles (ABPs) \cite{henkesDenseActiveMatter2020}, with constant speed $v_0$ in a direction set by the unit-length vector $\hat{\vb{n}}$ that follows rotational diffusion and is uncorrelated between cells.
We extended the original AVM \cite{bartonActiveVertexModel2017} to include internal dissipative effects in the form of pair-friction between cells, as a form of local viscosity. The importance of such effects was recently highlighted \cite{tong2023linear,rozmanSubstrateDissipationInternal2024}, and we show below that they are essential to reproduce the experimental spatial velocity correlations. Altogether, the resulting overdamped equation of motion is a force balance between dissipative, mechanical, and active forces, and reads
\begin{subequations}
\label{eq:AVM_eom}
\begin{eqnarray}
    \zeta \dot{\vb{r}}_i + \sum_{\langle ij \rangle} \zeta_{\mathrm{pair}} (\dot{\vb{r}}_i-\dot{\vb{r}}_j)&=& -\nabla_{\vb{r}_i} E_{\mathrm{AVM}} + \zeta v_0 \hat{\vb{n}}_i, \label{eq:motion-r} \\
    \dot{\hat{\vb{n}}}_i&=& \sqrt{\frac{2}{\tau}} \, \eta_i \, \hat{\vb{z}}\times\hat{\vb{n}}_i \label{eq:motion-n}    
\end{eqnarray}    
\end{subequations}
where $\dot{\vb{r}}_i(t)\equiv\vb{v}_i(t)$ is the velocity of cell $i$. In Eq.\ \eqref{eq:motion-r}, the first term on the left-hand side describes friction with a substrate, while the second term captures internal dissipation from cells moving past each other with different velocities; the sum is over the neighbours of cell $i$, including boundary points. Eq.\ \eqref{eq:motion-n} describes rotational diffusion of $\hat{\vb{n}}_i$ with persistence time $\tau$, where $\langle\eta_i(t)\rangle=0$ and \mbox{$\langle\eta_i(t)\eta_j(0)\rangle=\delta_{ij}\, \delta(t)$}, with $\langle\cdot\rangle$ denoting the ensemble average over realisations of the Gaussian white noise $\eta_i$. Here and in what follows, an argument $0$ in either the space or time coordinate within the average $\langle\cdot\rangle$ denotes a steady-state ensemble average over the origin of space or time, respectively. A graphical depiction of the model described by Eqs.\ \eqref{eq:AVM_eom} is shown in Fig.~\ref{fig:overview}e.

\subsection*{\label{sec:continuum_model}Origin of the bulk velocity correlations}
To understand velocity correlations and reproduce them in the AVM, we used a model of the cell monolayer based on dense active matter \cite{caprini2020hidden,mandal2020extreme,keta2022disordered}, and extended an existing continuum model \cite{henkesDenseActiveMatter2020} with viscoelasticity \cite{szamel2021long}. The overdamped equation of motion for the displacement field $\vb{u}(\vb{r},t)$ is represented as a continuous elastic solid (SI Appendix, section \ref{vcor}),
\begin{equation}
\begin{aligned}
    \zeta_c \dot{\vb{u}} = &B \nabla (\nabla \cdot \vb{u})  + \mu \nabla^2 \vb{u} + \zeta_c v_0 \vb{\hat{n}}  \\
    & + K \nabla (\nabla \cdot \dot{\vb{u}}) + \eta \nabla^2 \dot{\vb{u}}, 
\end{aligned}
\label{eq:continuum_eom}
\end{equation}
where we identify the velocity $\vb{v}(\vb{r}, t) \equiv \dot{\vb{u}}(\vb{r}, t)$.
Here, $\zeta_c$ is the substrate friction coefficient, and the monolayer is assumed to have elastic bulk and shear moduli $B$ and $\mu$. Similar to the AVM, we assume that the cells move independently as simple ABPs. In continuum, this takes the form of an active driving field $\zeta_c v_0 \vb{\hat{n}}(\vb{r},t)$, which is uncorrelated in space with a cutoff $a$ at the cell scale, but persistent in time with persistence time $\tau$. The vector field $\hat{\vb{n}}(\vb{r},t)$ follows the same stochastic dynamics as described by Eq.\ \eqref{eq:motion-n}.  
In addition, we now also include viscous terms: a bulk viscosity $K$ and a shear viscosity $\eta$ that correspond to the pair friction coefficient $\zeta_{\text{pair}}$ in the AVM. As we show below, they are essential to reproduce the experimental spatial and temporal velocity correlations. 

Analysing Eq.\ \eqref{eq:continuum_eom} in spatio-temporal Fourier space \cite{henkesDenseActiveMatter2020} (SI Appendix, section \ref{2d_velocity_field_derivation}) decouples the equation into longitudinal and transverse modes. Using the Lorentzian correlations of the driving field and then transforming back to real space, we obtain the 
spatial (\idest{}\ equal time) velocity correlation function for the viscoelastic isotropic continuum model,

\begin{equation}
\begin{aligned}
    &\left \langle \vb{v}(\vb{r}, 0) \vdot \vb{v}(\vb{0}, 0) \right \rangle = \frac{v_0^2 a^2}{4 \pi}  \frac{K_0(r/\xi_{\parallel,p}) - K_0(r/\xi_{\parallel,d})}{\xi_{\parallel,p}^2 - \xi_{\parallel,d}^2}  \\
    & + \frac{v_0^2 a^2}{4 \pi} \frac{K_0(r/\xi_{\perp,p}) - K_0(r/\xi_{\perp,d})}{\xi_{\perp,p}^2 - \xi_{\perp,d}^2},
\end{aligned}
\label{eq:continuum_spatial_vcor}
\end{equation}
where $r\equiv|\vb{r}|$ is the magnitude of vector $\vb{r}$, and $K_0$ denotes the $0^\mathrm{th}$ modified Bessel function of the second kind. 
This decaying correlation function is composed of two contributions. The first is from the longitudinal modes, with characteristic length scales \mbox{$\xi_{\parallel,p} = \sqrt{\left[(B + \mu)\tau + (K + \eta)\right]/\zeta_c}$} from the persistent driving and \mbox{$\xi_{\parallel,d} = \sqrt{\left(K + \eta\right)/\zeta_c}$} from the substrate and internal dissipation.
The second contribution is from the transverse modes, also having two characteristic length scales
\mbox{$\xi_{\perp,p} = \sqrt{\left(\mu\tau + \eta\right)/\zeta_c}$} from the driving and \mbox{$\xi_{\perp,d} = \sqrt{\eta/\zeta_c}$} from the dissipation.
By comparing Eq.\ \eqref{eq:continuum_spatial_vcor} and Ref.~\cite{henkesDenseActiveMatter2020} we see that including bulk and shear viscosities $K$ and $\eta$ leads to the two extra hydrodynamic length scales $\xi_{\parallel, d}$ and $\xi_{\perp, d}$, while modifying the two original length scales $\xi_{\parallel, p}$ and $\xi_{\perp, p}$ of active elastic origin which retain their dominant $\sqrt{\tau}$ scaling.

We note that $K_0(r/\xi) \sim -\log (r/\xi)$ in the limit $r \to 0$, which diverges.  The differences as in Eq.\  \eqref{eq:continuum_spatial_vcor}, however, converge to a finite value for $r \to 0$.
Therefore, the role of internal dissipative effects is to regularise short-range correlations, which is instrumental to reproducing the smooth experimental velocity correlations. Likewise, AVM simulations without pair dissipation show large velocity fluctuations between neighbouring cells \cite{biMotilityDrivenGlassJamming2016,henkesDenseActiveMatter2020}.

In summary, the effect of the elastic and viscous interactions between the cells is to turn the spatially \textit{uncorrelated}, but persistent, driving from the crawling cells into a spatially \textit{correlated} velocity field. 

\begin{table*}[ht]
    \centering
    \renewcommand{\arraystretch}{1.15}   
    \begin{tabular}{lccc|p{4.3cm}}
        \hline
        \textbf{Free parameter} & \textbf{Symbol} & \textbf{Value} & \textbf{Unit} & \textbf{Experimental measure} \\ \hline
        crawling persistence time        & $\tau$                    & 1.0 $\pm$ 0.5      & h                                            & timescale of velocity autocorrelation \\ 
        crawling speed                   & $v_0$                     & 255 $\pm$ 20    & \si{\micro\meter\per\hour}                   & root mean square velocity \\ 
        pair-friction coefficient        & $\zeta_{\text{pair}}/\zeta$ & 2.0 $\pm$ 0.5      & —                                            & initial decay spatial velocity correlation\\
        perimeter modulus                & $K_P/\zeta$               & 14 $\pm$ 2     & h$^{-1}$                                     & decay length spatial velocity correlation \\ 
        stretching spring constant           & $k_s/\zeta$               & 78 $\pm$ 50      & h$^{-1}$                                     & shape of tangent-tangent correlation \\ \hline
        \textbf{Fixed parameter} &  &  &  &  \\
        target area                      & $A_0$                     & 171    & \si{\micro\meter\squared}                    & mean cell area \\

        target shape index               & $p_0 = P_0/\sqrt{A_0}$    & 3.72   & —                                            & hexagonal ground state \\ 
        bending spring constant          & $k_b/\zeta$               & 1700   & \si{\micro\meter\squared\per\hour}           & numerical stability (AVM) \\ 
        area modulus                     & $K_A/\zeta$               & $K_P/(\zeta A_0)$ & \si{\per\micro\meter\squared\per\hour} & dimensional scaling \\ \hline
    \end{tabular}
    \vspace{1em}
    \caption{Parameters of the non-dividing Active Vertex Model and corresponding experimental constraints. Please refer to the main text for a detailed discussion of the calibration process of the free parameters. The interval around the matched parameters denotes the distance to the next parameter value(s) considered in the calibration process, rounded to one significant digit.}
    \label{tab:free_parameters}
\end{table*}

\subsection*{\label{sec:comparison_to_experiments}Comparison to experiments}
We are able to quantitatively match the experimentally observed spatiotemporal dynamics of the bulk of the monolayer to the AVM and the continuum viscoelastic model using an iterative process consisting of the following steps, successively refined (Tab.\ \ref{tab:free_parameters}).

We determined AVM target area and unit of length $\sqrt{A_0}$ by segmenting the cell shapes (SI Appendix, section \ref{segmentation}) and computing the average cell area (SI Appendix, Fig.\ \ref{fig:area_perimeter_shape_index_hist}). We set the target shape index to $p_0 = 3.72$ to ensure a solid ground state \cite{biMotilityDrivenGlassJamming2016}, while checking that the average area, perimeter, and shape index of the cells in the simulations matched the experiment (SI Appendix, Fig.\ \ref{fig:area_perimeter_shape_index_evolution}). To reduce the total number of parameters, we used a single (perimeter) stiffness $K_P$ for the AVM and determined the area stiffness $K_A = K_P/A_0$ from dimensional scaling (Tab.\ \ref{tab:free_parameters}). 
Subsequently, we inferred parameters for the continuum elastic model by simple dimensional mapping from the AVM parameters (SI Appendix, section \ref{AVM_interior_mapping}).

The temporal velocity autocorrelation function \mbox{$C_{\mathrm{\vb{v}\vb{v}}}^{\mathrm{T}}(t)=\langle \vb{v}^*(\vb{0},t) \cdot \vb{v}^*(\vb{0},0) \rangle$}, where $\vb{v}^*(\vb{r},t) =\vb{v}(\vb{r},t)/v_{\text{rms}}(t)$ is the velocity vector normalized by the root mean square velocity at time $t$, measures the persistence time of the individual velocity vectors of the cells (Fig.\ \ref{fig:velocity_statistics_panel}a, experiment in black). 
We observe two regimes in the experiments: a sharp decay for approximately 2 \si{\hour} followed by a much gentler decay.
The autocorrelation function of the AVM is only strongly influenced by the persistence time $\tau$ of the driving. We found a very good match to the first regime of the experiments for $\tau\!=\!1$ h (Fig.\ \ref{fig:velocity_statistics_panel}a and b, green curves). Similarly, the theoretical autocorrelation function is dominated by the exponential factor $\sim\!e^{-t/\tau}$ (SI Appendix, section \ref{sec:cvvt}), and we show the numerically integrated prediction for $\tau\!=\!1$ (Fig.\ \ref{fig:velocity_statistics_panel}a and b,  orange curves).

We pair this with the spatial velocity correlation function, which measures the spatial extent of the correlated cell movements, \mbox{$C_{\vb{v}\vb{v}}^{\mathrm{S}}(\vb{r})=\langle \vb{v}^*(\vb{r},0) \cdot \vb{v}^*(\vb{0},0)\rangle$} (Fig.\ \ref{fig:velocity_statistics_panel}b, experiment in black), where because of isotropy we assume $C_{\vb{v}\vb{v}}^{\mathrm{S}}(\vb{r}) = C_{\vb{v}\vb{v}}^{\mathrm{S}}(r)$ with $r\!=\!|\vb{r}|$. It shows exponential correlations of cell migration over a range of $\xi\! \sim\!100$ \si{\micro\meter} (i.e.\ of the order of 10 cell widths), consistent with previous observations in MDCK monolayers \cite{petitjeanVelocityFieldsCollectively2010,garciaPhysicsActiveJamming2015,henkesDenseActiveMatter2020}.
In contrast to others \cite{serra-picamalMechanicalWavesTissue2012,boocock2021theory}, we did not observe space-time correlations like wave propagation. Instead, space and time correlations remain decoupled (see kymographs in SI Appendix, section \ref{kymographs}). 

After varying the perimeter stiffness $K_P$ and the pair-friction coefficient $\zeta_{\mathrm{pair}}$, we obtained an excellent match using the AVM (Fig.\ \ref{fig:velocity_statistics_panel}b, in green). This also matches the corresponding continuum theory result,   Eq.\ \eqref{eq:continuum_spatial_vcor} (Fig.\ \ref{fig:velocity_statistics_panel}b, in orange), again computed using dimensional scaling only. In contrast, the theory result for the case without any internal dissipation (in purple) leads to a large drop in the spatial velocity correlation at small length scales. Despite this initial drop, the function regains its exponential character at longer length scales (Fig.\ \ref{fig:velocity_statistics_panel}b, inset).

According to our linear theory (Eq.\ \eqref{eq:continuum_spatial_vcor} and SI Appendix, sections \ref{sec:cvvt} and \ref{sec:cvvr}), the normalised correlation functions do not depend on the driving amplitude $v_0$ but, as we show below, finger amplitudes do. Therefore, to constrain $v_0$ to realistic values, we aim to match the root mean square velocity $v_{\mathrm{rms}}$ of the cells in the finger region.
These cell velocity decreases slowly over time (Fig.\ \ref{fig:velocity_statistics_panel}a), a trend qualitatively similar to what was previously found  \cite{petitjeanVelocityFieldsCollectively2010}. For $v_0\!=\!255$ \si{\micro \meter\per\hour}, we find for the AVM that, after a quick relaxation to steady state,  $v_{\mathrm{rms}}\!\approx\!22$ \si{\micro \meter\per\hour}, which is a good match to the experiments. The large difference between $v_0$ and $v_{\mathrm{rms}}$ indicates that the elastic and dissipative cell-cell interactions, and thus the confining effects of neighbouring cells, are very strong.

\section*{\label{sec:finger_formation}Results - finger formation}
\begin{figure*}
    \centering
    \includegraphics[width=0.9\textwidth]{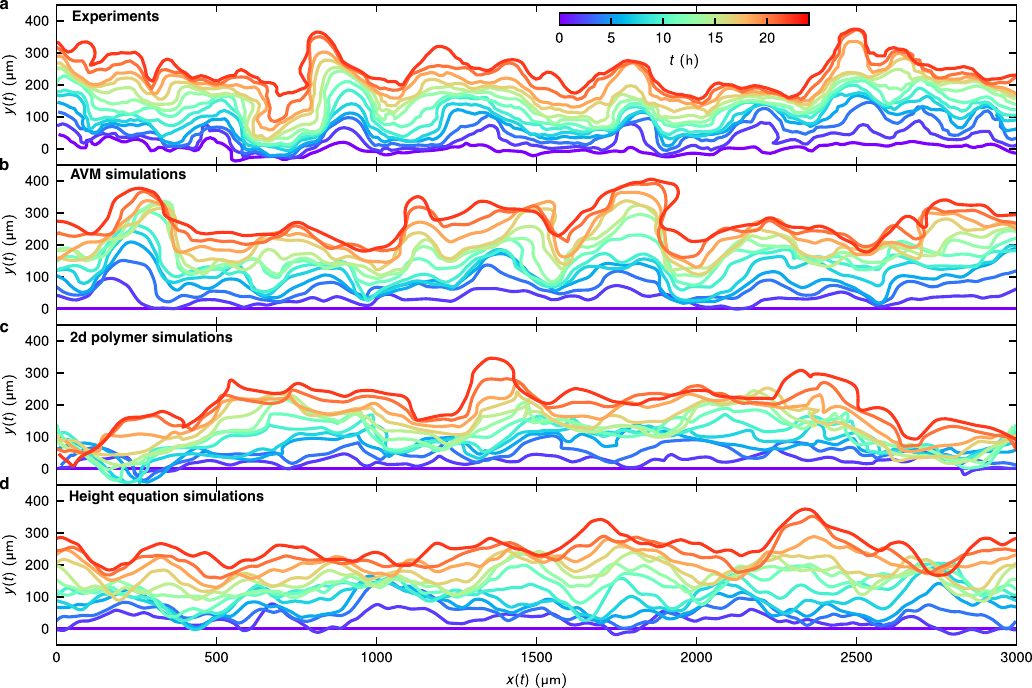}
    \caption{Example boundary contours at different times of (\textbf{a}) experiments, (\textbf{b}) AVM simulations, (\textbf{c}) active polymer simulations and (\textbf{d}) height equation simulations. Panels \textbf{b-d} have been shifted by the experimental mean border progression, as discussed in the main text. }
    \label{fig:contour_panel}
\end{figure*}

Having established that the AVM and continuum theory can both reproduce the experimentally observed velocity correlations in the bulk, we now focus on the behaviour of the advancing edge of the monolayer in the experiments and the AVM.
To do so, we extracted the contour of the advancing edge as a function of time from experiments (Fig.\ \ref{fig:overview} and SI Appendix, sections \ref{image_processing} and \ref{spline_smoothing_grid_binning}). 
Consistent with other experimental results \cite{poujadeCollectiveMigrationEpithelial2007,vishwakarmaMechanicalInteractionsFollowers2018}, the fingers developed over a few hours after the insert 
has been lifted, and cells were allowed to spread (Movies \ref{mov:experiment_inverted_grayscale} and \ref{mov:experiment_analyzed}). The fingers were persistent, i.e.\ they stayed in place relative to the mean border position for many hours (Fig.\ \ref{fig:contour_panel}a), although we did observe occasional retractions. We observed a wide variety in the width and amplitude of the fingers.

In the AVM, for sufficiently small values of the spring constant $k_s$, we obtained similar finger formation (Movies \ref{mov:non_expanding_AVM_full} and \ref{mov:non_expanding_AVM_zoom}, and Fig.\ \ref{fig:contour_panel}b). Since in the absence of cell division the AVM cell sheet does not expand, the AVM contours in Fig.\ \ref{fig:contour_panel}b are displaced by the mean position of the advancing edge in experiments. We will show later that including cell division and flattening in the AVM reproduces the average border progression observed in the experiments. The AVM simulations also showed the long lifetime of fingers and the variety of finger sizes. 

We now compare simulated and measured advancing edges quantitatively by defining three boundary correlation functions, which allow us to constrain the remaining parameters of the AVM, the boundary spring constant $k_s$ and refine the crawling speed $v_0$.

\subsection*{\label{sec:finger_formation_reaches_steady_state}Fingers have different sizes and reach steady state}
\begin{figure*}
    \centering
    \includegraphics[width=0.75\textwidth]{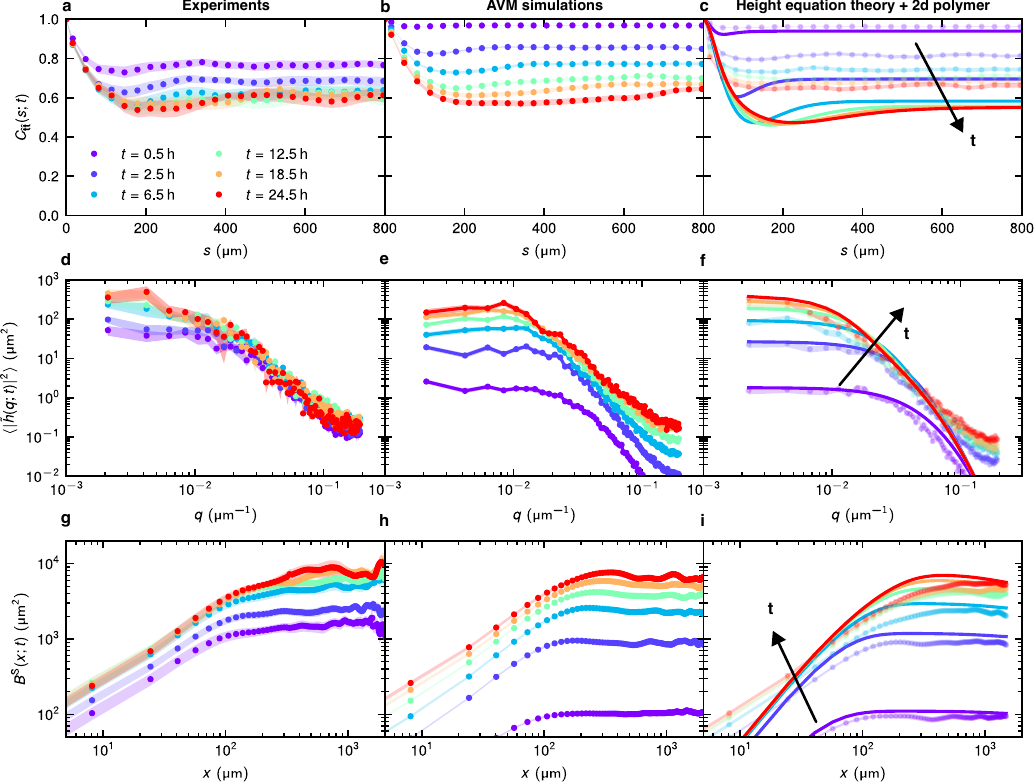}
    \caption{Spatial properties of the fingers as a function of time. The left column corresponds to experiments, the middle column to AVM simulations, and the right column to polymer simulations (dots) and height equation theory (lines). The shading indicates the standard error of the mean. (\textbf{a-c}) The tangent-tangent correlation function measures the correlation of the slope of the boundary as a function of distance $s$ along the boundary over time. (\textbf{d-f}) Spatial Fourier spectrum of the height fluctuations over time. (\textbf{g-i}) The relative roughness correlation function measures the roughness of the height on a length scale $x$ over time. }
    \label{fig:tangent_spatial_roughness_panel}
\end{figure*}

We define the tangent-tangent correlation function, commonly used for polymers \cite{doi1988theory,golestanian2000statistical,liverpool2001dynamic,liverpoolAnomalousFluctuationsActive2003},
\begin{equation}
   C_{\hat{\vb{t}}\hat{\vb{t}}}(s ; t)\equiv \left \langle  \hat{\vb{t}}(s,t) \cdot \hat{\vb{t}}(0,t)\right \rangle,
\end{equation}
where $s$ is the arc-length coordinate of the curve describing the advancing edge, $t$ is the time since the initial state, and $\hat{\vb{t}}(s, t)$ is the local unit-length tangent vector. 
The tangent-tangent correlation measures the correlation of the slopes of any two points along the curve, separated by $s$, and we explicitly track its evolution through time $t$. 
In experiments (Fig.\ \ref{fig:tangent_spatial_roughness_panel}a), as the fingers form, there are no strong oscillations along $s$, and thus no sign of a finite-wavelength instability. Instead, the tangent-tangent correlation shows a strong decay over the first 200 \si{\micro\meter}, a length scale which we identify with finger formation and which is of the same order as the velocity correlations (Fig.\ \ref{fig:velocity_statistics_panel}b). After this, $C_{\hat{\vb{t}}\hat{\vb{t}}}(s;t)$
reaches a plateau at large separation, with a steady state value around 0.6 at large times.

In AVM simulations, the tangent-tangent correlation 
shows the same qualitative trend as in the experiments (Fig.\ \ref{fig:contour_panel}b). After matching the edge stretching stiffness $k_s$ and refining $v_0$ (constrained by $v_{\text{rms}}$) to match the amplitude of the temporal decay, it reaches the same steady state value, although not as quickly as in the experiments (Fig.\ \ref{fig:tangent_spatial_roughness_panel}b). The stronger decay at $t\!=\!\SI{0.5}{\hour}$ in the experiments compared to the AVM simulations is caused by the advancing edge in experiments not being completely flat at the start of the measurement, unlike in simulations (Fig.\ \ref{fig:contour_panel}a and SI Appendix, Fig.\ \ref{fig:suitable_series}, see  Fig.\ \ref{fig:polymer_sims_initial_roughness} for simulations starting from a non-flat edge).
All the free parameters and the corresponding experimental constraints are summarised in Tab.\ \ref{tab:free_parameters}.

We further characterise the boundary properties by mapping to a one-dimensional height interface.
The observed contours sometimes have overhangs, and the horizontal spacing between points is irregular (Fig.~\mbox{\ref{fig:contour_panel}a and b}). We, therefore, binned the $y$-coordinates of all points on a regular grid of $x$-coordinates by taking the maximum values of $y$ in the bin (SI Appendix, section \ref{spline_smoothing_grid_binning}), thereby obtaining the height function $h(x)$ of the boundary.  We note that this method has been used before in the study of cell sheet borders \cite{rapinRoughnessDynamicsProliferating2021}.

A finite-wavelength instability should show a prominent peak in the spatial Fourier spectrum of the boundary height $\langle| \tilde{h}(q;t)|^2 \rangle$. 
However, as the fingers form on the boundary, we see no peak appearing in either the experiments or AVM simulations (Fig.\ \ref{fig:tangent_spatial_roughness_panel}d and e). Instead, the low $q$ fluctuations gradually grow over time until they converge after $\approx\!24$ hours.
This approach is more pronounced in the AVM as it starts from a perfectly flat border. We also see a change towards an approximate $q^{-2}$ slope around, once again, a lengthscale of $2 \pi q^{-1} \approx \SI{100}{\micro \meter}$.
We will return to the noticeable differences at large wavenumber $q$ when we formulate a height equation theory for the finger formation.

Another probe for the spatial properties of the advancing edge is the relative roughness function \cite{agoritsas2012disordered}, previously used to study other cell fronts \cite{rapinRoughnessDynamicsProliferating2021}. For a boundary contour represented by a vertical height $h$ along a horizontal direction $x$ (Fig.\ \ref{fig:contour_panel}), it is defined by
\begin{equation}
    B^{\mathrm{S}}(x; t) = \left \langle  \left[ h(x,t) - h(0,t)\right]^2 \right \rangle,
    \label{eq:relative_roughness}
\end{equation}
and measures the relative roughness of points on the boundary with horizontal spacing $x$ at time $t$ since the initial state; $\langle\cdot\rangle$ denotes the average over all conformations of the boundary contour. 

Roughness in both experiments and AVM simulations reached a steady state at all scales (Fig.\ \ref{fig:tangent_spatial_roughness_panel}d and e). Furthermore, we again did not observe peaks or oscillations, consistent with a variety of finger sizes. Like with the tangent-tangent correlation, a crossover occurs at $\sim\!\SI{100}{\micro \meter}$ after which the relative roughness increases and stabilises. Furthermore our results are consistent with the experimental data in \cite{rapinRoughnessDynamicsProliferating2021}.

\subsection*{\label{sec:finger_formation_persistent}Fingers have a long lifetime}
\begin{figure*}
    \centering
    \includegraphics[width=1\textwidth]{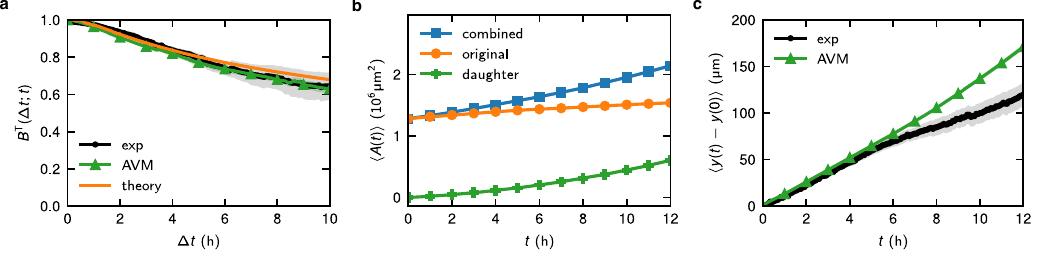}
    \caption{Temporal properties of the boundary. The shading indicates the standard error of the mean. (\textbf{a}) The height autocorrelation function measures the correlation between the boundary shape at time $t$ and the shape at $t+\Delta t$, shown here for $t=\SI{24}{\hour}$ for experiments (black dots), the non-expanding AVM simulations (green triangles) and the height equation theory (orange line). (\textbf{b}) Average cell area over time in the expanding AVM simulations, specified for the initial cells (orange dots), daughter cells (green plusses) and the total (blue squares). (\textbf{c}) Average progression of the cell sheet boundary over time, \idest{}\ the average $y$ coordinate over time, in experiments (black dots) and expanding AVM simulations (green triangles). For clarity, only every tenth AVM marker is shown.}
    \label{fig:temporal_expansion_panel}
\end{figure*}

We quantify the persistence time of fingers by defining the height autocorrelation function
\begin{equation}
    B^{\mathrm{T}}(\Delta t; t) \equiv \frac{1}{N} \left \langle \Delta h(0,t+\Delta t) \Delta h(0,t)\right \rangle,
    \end{equation}
in which $\Delta h(x,t)\equiv h(x,t) - \overline{h}(t)$ is the deviation from the mean boundary height of the $h$-coordinates at time $t$. This function can be understood graphically as the average point-wise product of the $h$-coordinates of the contour at time $t$ (\mbox{Fig.~\ref{fig:contour_panel}a-c}) and a contour $\Delta t$ later, normalised by the standard deviations at the respective times, $N= \sqrt{\left \langle \Delta h(0,t+\Delta t)^2\right \rangle} \sqrt{\left \langle \Delta h(0,t)^2\right \rangle}$.
We show $B^{\mathrm{T}}(\Delta t; t)$ for experiments and AVM simulations in \mbox{Fig.~\ref{fig:temporal_expansion_panel}a} for $t=\SI{24}{\hour}$, when the finger formation has reached steady state.

The boundary contours and thus the fingers in both the experiments and the AVM simulations are long-lived, retaining their shape over multiple hours (Fig.\ \ref{fig:temporal_expansion_panel}a), which is much longer than the persistence time ($\SI{1}{\hour}$) of the cell crawling direction in the AVM simulations. We note that at early times $t$, the experiments show a longer persistence of the fingers than the AVM simulations (SI Appendix, Fig.\ \ref{fig:early_time_boundary_persistence}), which we attribute to the small pre-existing roughness at the start of the experimental imaging (Fig.\ \ref{fig:contour_panel} and SI Appendix, Fig.\ \ref{fig:suitable_series} and  Fig.\ \ref{fig:polymer_sims_initial_roughness}).

Together, the measurements of the tangent-tangent correlation $C_{\hat{\vb{t}}\hat{\vb{t}}}(s;t)$, the Fourier spectrum $\langle |\tilde{h}(q;  t)|^2\rangle$, and the spatial  $B^{\mathrm{S}}(x; t)$ and temporal $B^{\mathrm{T}}(\Delta t; t)$ roughness correlations corroborate the observation that fingers come in a variety of sizes, last a long time, and that the border progresses to steady state.

\subsection*{\label{sec:height_equation_theory}Polymer and height equation theory for finger formation} 
We seek to understand the origin of the fingers, and their properties as they 
arise from the interplay between the bulk velocity fluctuations and the contractile edge of the cell monolayer.
Motivated by the actomyosin cable linking cells and its boundary implementation in the AVM, we model the edge as a semiflexible polymer driven by the motion of the cells of the interior, with overdamped equations including pair dissipation
\begin{equation}
    \zeta \dot{\vb{r}}_i +\zeta_{\mathrm{pair}} (\dot{\vb{r}}_{i+1}+\dot{\vb{r}}_{i-1}-2\dot{\vb{r}}_i)= -\nabla_{\vb{r}_i}E_{\mathrm{edge}}+\zeta \vb{v}^f_i.
\end{equation}
Here $i$ labels points in the polymer with periodic boundary conditions with horizontal length $L_x\!=\!\SI{3000}{\micro\meter}$, matched to the experiment, and where $ \vb{v}^f_i$ is the driving by the velocity field of the monolayer with correlations given by Eq.\ \eqref{eq:continuum_spatial_vcor}. We observed an excellent match between the active polymer description and the AVM simulations (Fig.\ \ref{fig:contour_panel}c and Fig.\ \ref{fig:tangent_spatial_roughness_panel}c, f and i, and Movie \ref{mov:polymer}). This shows that a simple driven model with no feedback instabilities is able to quantitatively reproduce finger formation. Note, however, that we had to fit a very small value of spring constant $k_s$, likely reflecting the effectively viscoelastic relaxation mechanism of the cable in both experiment and AVM (SI Appendix, section \ref{mapping_AVM_boundary_to_heq_and_polymer}).

To gain insight into the origin of finger length and time scales, we derive an equivalent continuum formulation starting from a (stretched) worm-like chain~\cite{marko1995stretching,saarloosSoftMatterConcepts2024} extended to include correlated active driving. 
In the limit of small fluctuations and the absence of overhangs, the active polymer can be described by an equation of motion for its vertical height $h$ along a horizontal direction $x$ (SI Appendix, section \ref{heq_derivation})),
\begin{equation}
\begin{aligned}
    \zeta_h \partial_t h(x,t)={}&\lambda \frac{\partial^2 h(x,t)}{\partial x^2}  - \kappa \frac{\partial^4 h(x,t)}{\partial x^4} \\
    &+ \eta_h \frac{\partial^2 \partial_t h(x,t)}{\partial x^2}   + \zeta_h v^{f}(x,t),
\end{aligned}
\label{height equation, pair-dissipation}
\end{equation}
referred to below as the height equation, where we have assumed overdamped dynamics for the advancing edge with
friction coefficient $\zeta_h$. The line tension $\lambda$ and bending stiffness $\kappa$ in the worm-like chain contribute the first two terms. We also included pair-dissipation (viscosity $\eta_h$), like in \cite{khatri2007rouse}. The boundary is driven by the stochastic velocity field $v^{f}(x,t)$ from the cell movements in the interior, for which we derive an auxiliary one-dimensional equation based on the viscoelastic model (SI Appendix, section \ref{mapping_AVM_boundary_to_heq_and_polymer}). As with the two-dimensional continuum model for the velocity correlations, we can directly map the AVM parameters to the height equation parameters, except for again needing a much lower value for the line tension $\lambda$ (SI Appendix, section \ref{mapping_AVM_boundary_to_heq_and_polymer}).

Simulations of the height equation exhibit boundary fluctuations (Fig.\ \ref{fig:contour_panel}c and Movie \ref{mov:heq}) that capture the length scales of experiment and simulations and persist over multiple hours, though by construction, not overhangs.
The height equation analytically reproduces the tangent-tangent correlation of the experiment and simulations (Fig.\ \ref{fig:tangent_spatial_roughness_panel}c). We also observed a Fourier spectrum (Fig.\ \ref{fig:tangent_spatial_roughness_panel}f) and relative roughness correlation (Fig.\ \ref{fig:tangent_spatial_roughness_panel}i) that matches the AVM for length scales larger than \SI{100}{\micro \meter}.  

We can understand the dominant length and time scales through analysis of the correlations of the linear theory (SI Appendix, section \ref{sec:height-analysis}, see Eq. \eqref{hh, qt-space}). The tangent-tangent correlation contains five different pairs of poles on the imaginary axis at wave vectors $q_i$ (Eq. \eqref{eq:pole1-10}), setting five length scales $\xi_i = 2\pi/q_i$. 
The largest and dominant of these length scales is for $q_1 = \sqrt{\frac{\zeta}{\mu\tau + \eta}}$, the same as the shear length scale of the interior velocity correlations. 
In the overdamped one-dimensional dynamics of the edge, mode $q$ relaxes with a time scale $\tau_q \sim q^{-2}$. The finger lifetime is then dominated by relaxation at scale $q_1$, explaining their longevity. 
In the height correlator, there is a leading $1/q^2$ term that becomes dominant in the infinite time, infinite system size limit. The length scales (Fig.\ \ref{fig:tangent_spatial_roughness_panel}f and i) are thus coarsening over the course of the experiment, albeit very slowly. In contrast, the tangent-tangent correlations reach a true steady state since $\hat{\bm t} \sim \bm \nabla h$, and so do not contain the $1/q^2$ - in fact, their thermal limit is the strong force expansion of the WLC model \cite{saarloosSoftMatterConcepts2024}.

There is a notable difference between the height equation and the two-dimensional active polymer simulations (Fig.\ \ref{fig:tangent_spatial_roughness_panel}f and i). It emerges at short length scales (i.e.\ high values of the wavevector $q$) and increases with time, with the polymer but not the height equation matching the AVM and with the experiment resembling the longest times. 
We attribute this discrepancy to nonlinear effects stemming from the coupling, at large deformations, between height and transverse fluctuations along the stretchable polymer chain. It will be interesting to classify this nonlinearity, which does not appear to be the KPZ term \cite{KPZ1986} and may instead be effectively thermal \cite{caballero2022microscopic}. 
As the experiment does not start from a perfectly straight boundary (Fig.\ \ref{fig:contour_panel}a and SI Appendix, Fig.\ \ref{fig:suitable_series}), these effects appear immediately (see Fig.\ \ref{fig:polymer_sims_initial_roughness} for simulations with initial roughness). It is responsible for shifting the short-time roughness exponent from approximately $2$ to $1$, consistent with previous observations \cite{rapinRoughnessDynamicsProliferating2021}.

\section*{\label{sec:division_flattening}Cell proliferation and flattening}
The finger formation can be thought of as a roughening effect of the advancing edge in addition to the average expansion of the sheet. Previous experiments showed \cite{poujadeCollectiveMigrationEpithelial2007}  that inhibiting cell division did not abolish finger formation for $t<\SI{10}{\hour}$, indicating that proliferation is not the driving factor. 
This is consistent with our base model. However, it is important to determine whether the interface is affected by the effects of proliferation over longer time scales. 
To this end, we included stochastic cell division in the AVM. Cells grow by slowly increasing their preferred area $A_{i,0}(t)$ over time, with $A_{i,0}$ at time $t=\SI{0}{\hour}$ set to vary slightly between cells to prevent synchronised division (Methods). The division rate scales with $A_i(t)-A_{i,0}$ where $A_i(t)$ is the cell's actual area \cite{bellCellGrowthDivision1967,bartonActiveVertexModel2017}. This leads to a tissue with a uniform growth rate and spatially randomly distributed divisions. 

An additional process that has to be considered in this context is cell flattening, which occurs immediately after the barrier is released.
This response occurs right after the removal of the containment, during which the cells lose height and expand in area \cite{serra-picamalMechanicalWavesTissue2012}. 
We model it in the AVM by changing the initial dimensions of the cell strip, so that the initial average cell area was \mbox{$\overline{A}(t=0) = 0.90 A_0$}. This puts the tissue initially in a slightly compressed state, consistent with observations in tissues and cultured epithelial cells \cite{eisenhoffer2013bringing,zulueta2022role}, causing the initial cells to expand in the first few hours after barrier removal.

The expanding AVM (Movie \ref{mov:expanding_AVM}) matches well to the experimental mean border progression for the first $\SI{5}{\hour}$ (Fig.\ \ref{fig:temporal_expansion_panel}c), with contributions from both cell expansion and division (Fig. \ref{fig:temporal_expansion_panel}b). It then starts to deviate, as after this time, the experiments reveal decreased border progression, which we do not reproduce. A more detailed model of cell division and spreading is needed to capture this crossover. Importantly, the properties of the interface (tangent-tangent correlations, spectrum and roughness) remain quantitatively consistent with the simulations without division (SI Appendix, Fig.\ \ref{fig:tangent_roughness_spectrum_expanding_AVM}).

The addition of the two mechanisms also produces a velocity gradient towards the edge of the tissue in the AVM (SI Appendix, section \ref{kymographs} ).
This causes the temporal velocity autocorrelation in \ref{fig:velocity_statistics_panel}a  to develop a heavier tail at long times, a trend similar to that observed in the experiments (SI Appendix, Fig.\ \ref{fig:autocorrelation_expanding_AVM}).

\section*{\label{sec:conclusion}Discussion and Conclusion}

Here we have shown how correlated cell movements can drive finger formation at the boundary of cell sheets.
Using experiments on epithelial MDCK monolayers, as well as simulations of the Active Vertex Model (AVM) calibrated to reproduce their collective behaviour, we showed that finger formation on the boundary reaches a steady state and leads to long-lived fingers of a variety of sizes.

Our results do not support the existence of a finite-wavelength instability in finger formation. Instead, we reproduced experiments and simulations quantitatively by combining the space and time correlated collective cell motion in the interior with the contractile effect of the multicellular contractile actomyosin cable surrounding the tissue. 
The characteristic length and time scales of the fingers are controlled by the stiffness of the tissue and the contractility of the actomyosin cable, as well as by the persistence time of active cell crawling.

It puts our results into a similar class of active interfaces as the active nematic filaments of \cite{adkins2022dynamics,zhao2024asymmetric}, but it remains to be seen how it relates to other active interface problems \cite{cates2025active}. 

Our model makes the experimental prediction that finger length and time scales are directly set by the spatiotemporal correlations of the interior and the actomyosin cable, and their growth rate by the activity. As has been highlighted before, the internal correlation length is sensitive to density \cite{garciaPhysicsActiveJamming2015} and changes in the substrate stiffness \cite{vazquez2022effect}, in the latter case also directly affecting the finger length scale. In our fitted model, the hydrodynamic length scale $\xi_{\perp,d} = \sqrt{\eta/\zeta_c}$ and which has been linked to the correlation length \cite{garciaPhysicsActiveJamming2015,vazquez2022effect}, is only approximately two cell diameters. Ultimately, we predict that modifications that change either cell-cell adhesion $K_P$, either friction coefficient, crawling motility $v_0$, or the properties of the actomyosin cable $\lambda$ will changes both interior and finger length and time scales according to our theory, dominated by the active elastic length scale $\xi_{\perp,p} = \sqrt{(\mu \tau + \eta)/\zeta_c}$.

At this point, the question arises how our results connect to the many feedback mechanisms that have been observed experimentally (notably leader cells \cite{vishwakarmaMechanicalInteractionsFollowers2018} and plithotaxis), and proposed theoretically (\exgr{}\ curvature enhanced motility \cite{markPhysicalModelDynamic2010}). Our model can be considered as a fluctuation-driven \emph{null model}, \idest{}\ the basic phenomenology arising from \emph{only} uncorrelated motility and viscoelastic mechanics. The critical conclusion is that an unexpectedly large proportion of experimental observations are well explained by such a simple model. We hypothesise that additional feedback, acting on the background of the uncorrelated driving, is responsible for the \emph{details} of the phenomenology observed. For example, leader cells arise at the tip of forming fingers from the local mechanical environment \cite{vishwakarmaMechanicalInteractionsFollowers2018} and may then amplify their formation.

\section*{Methods}

\subsection*{Experiments} Madin-Darby canine kidney (MDCK) cells were seeded into Ibidi two-well culture inserts \cite{ibidiCultureInsertWell} which had been attached to TTP 40mm $\oslash$ Tissue Culture Petri Dishes (93040T). Each well had a 0.22 $\si{\cm\squared}$ surface area. A 70 $\si{\micro\liter}$ cell suspension was added to each so that the cells were fully confluent after an overnight incubation. To initiate cell movement and create an empty space, \idest{}\ ``wound'', the insert was removed, leaving cells free to spread (Fig.\ \ref{fig:overview}a). To record the moving edge, 5 overlapping fields of view along each free edge were imaged every 5 minutes for 72 $\si{\hour}$ – this was to ensure the cells could be observed for the maximum time and had reached and gone past the edge of the field of view.  Consequently, the analysis was over a shorter period (usually 40 $\si{\hour}$) and included only the time points when the edges were in field of view. The 5 fields of view were stitched together post-acquisition.  The recording microscope was a Nikon TiE with a Nikon Plan Fluor ELWD 20x NA0.45 Ph1 ADL objective (using phase contrast). Images were recorded with Photometrics Cascade II 1K camera. Heating to 37$\si{\celsius}$ and 5\% $\mathrm{CO_2}$ during the recording were provided by an Okolab incubation system.

\subsection*{AVM simulations}
We extended the Active Vertex Model (AVM) \cite{bartonActiveVertexModel2017}, a variation of the self-propelled Voronoi model \cite{liCoherentMotionsConfluent2014,biMotilityDrivenGlassJamming2016} with a dynamic cell sheet boundary. 
This model represents a confluent epithelial cell monolayer by a two-dimensional polygonal Voronoi tiling, with cell centres located on the vertices of the dual Delaunay triangulation, and it is part of the wider class of vertex models (VM) \cite{nagaiDynamicCellModel2001,farhadifarInfluenceCellMechanics2007,fletcherVertexModelsEpithelial2014}.
Cells in the interior experience the standard vertex model potential
\begin{equation}
    E_{\mathrm{VM}} = \sum_{\mathrm{C}=1}^{N_{\mathrm{C}}} \left[ \frac{K_A}{2} (A_{\mathrm{C}} - A_0)^2 + \frac{K_P}{2}(P_{\mathrm{C}}- P_0)^2\right], \label{eq:VM}
\end{equation}
 modelling elastic cell-cell interactions with area modulus $K_A$ (cell volume incompressibility) and perimeter modulus $K_P$ (cell-cell adhesion and cell contractility), as well as target (i.e.\ preferred) area $A_0$ and target (i.e.\ preferred) perimeter $P_0$. As in the original AVM \cite{bartonActiveVertexModel2017}, we add a short-range soft core repulsive potential to avoid problems with highly obtuse triangles in the dual Delaunay triangulation.

The advancing edge of the cell monolayer acts as a semi-flexible polymer under contraction, surrounding the tissue and closing the finite Voronoi tesselation. Its energy is 
\begin{equation}
    E_{\mathrm{edge}} = \sum_{i} \frac{k_s}{2}\left|\vb{r}_{i+1}-\vb{r}_i\right|^2 + \sum_{i} \frac{k_b}{2}(\theta_i - \pi)^2, \label{eq:Eedge}
\end{equation}
where index $i$ labels all points on the advancing edge, $a$ is the rest length of the bond connecting two consecutive boundary points, and $\theta_i$ is the angle between the two bonds meeting at the boundary point $i$. $k_s$ and $k_b$ are corresponding stiffnesses (i.e.\ spring constants). The inclusion of the angular springs is motivated by numerical stability, and $k_b$ is taken to be a constant sufficiently strong to prevent problems with the dual Delaunay triangulation. 

To include the dynamics into the model, we assume that cells crawl independently as simple Active Brownian Particles (ABPs) \cite{henkesDenseActiveMatter2020}. Model cells crawl at a constant speed $v_0$ in a direction that follows rotational diffusion
 i.e. $\vb{F}_i^{\mathrm{act}} =  \zeta v_0 \hat{\vb{n}}_i$ in which $\zeta$ is the substrate friction coefficient and $\vb{\hat{n}}_i = \cos(\phi_i)\hat{\vb{x}} +\sin(\phi_i)\hat{\vb{y}} $ is the polarity vector of cell $i$. Angle $\phi_i$ follows rotational diffusion, $\dot{\phi_i} = \sqrt{2/\tau} \, \eta_i$, where $\eta_i$ is a Gaussian white noise with zero mean $\langle \eta_i(t) \rangle =0$ and unit variance \mbox{$\langle \eta_i(t) \eta_j(0) \rangle = \delta_{ij} \, \delta(t)$}, where $\tau$ is the persistence time.

 We extended the original AVM to include internal dissipative effects, resulting in the equation of motion \eqref{eq:AVM_eom}. The dynamics of the cell sheet are obtained by integrating over time using the Euler-Maruyama method. The internal dissipation mixes velocities on the left-hand side in Eq.\ \eqref{eq:motion-r}, which requires solving a linear system of equations in the form $\sum_j M_{ij}\vb{v}_j=\vb{F}_i$ in each time step. The symmetric matrix $M_{ij}$ is sparse, and the system can be efficiently solved using the Eigen3 library \cite{eigenweb}.  
  The boundary points follow the same equation of motion but without self-propulsion, and the boundary can dynamically grow and shrink, making the boundary effectively viscoelastic \cite{bartonActiveVertexModel2017}.


The initial condition for the non-dividing AVM simulations is a rectangular box of $3400\times650$ $\si{\micro \meter\squared}$. We place 12,861 cells and 1,620 boundary points so that the average area of the cells is \SI{171}{\micro \meter\squared}. The boundary points are placed along the edge of the box with a spacing of \SI{5}{\micro \meter}. The cells are placed randomly inside the box and then relaxed by using Brownian dynamics with a soft-core repulsive potential to set the initial Voronoi seeds. The subsequent AVM simulation is run for 40 hours with a time step of 0.0001 h. The results for the non-dividing AVM are the result of 31 independent simulations with different seeds for both the initialisation and AVM simulation steps.

We ran 10 independent expanding AVM simulations of initial size $2000\times650$ $\si{\micro \meter\squared}$, containing initially 8399 cells and 1069 boundary points spaced $\SI{5}{\micro\meter}$, such that the initial cell area was on average $154$ \si{\micro \meter\squared}. The threshold area for each cell to divide was set to its initial, compressed, area.  The growth and division rates, as defined in \cite{bartonActiveVertexModel2017}, are set to match the experimental border progression (Fig.\ \ref{fig:temporal_expansion_panel}c).

\section*{Acknowledgments}
\articleacknow{{}
\section*{References}
\bibliography{references_v2}

\clearpage
\onecolumngrid
\appendix
\renewcommand\partname{Appendix}
\renewcommand\appendixname{}
\titleformat{\section}
    {\centering\bfseries\MakeUppercase}
    {\thesection.}
    {0.9em}
    {}
\titleformat{\subsection}
    {\centering\bfseries}
    {\thesubsection.}
    {0.9em}
    {}
\titleformat{\subsubsection}
    {\centering\itshape}
    {\thesubsubsection.}
    {0.9em}
    {}
\makeatletter
\def\thesection{\Roman{section}}
\def\thesubsection{\Roman{section}.\Alph{subsection}}
\def\p@subsection{}
\def\thesubsubsection{\Roman{section}.\Alph{subsection}.\arabic{subsubsection}}
\def\p@subsubsection{}
\makeatother
\counterwithout{figure}{section}
\counterwithout{table}{section}
\counterwithout{equation}{section}
\setcounter{figure}{0}
\setcounter{table}{0}
\setcounter{equation}{0}
\renewcommand\thefigure{S\arabic{figure}}
\renewcommand\thetable{S\arabic{table}}
\renewcommand\theequation{S\arabic{equation}}

\section*{Appendix table of contents}
\tableofcontents
\clearpage

\part{Supporting figures}
\section{Interior}

\subsection{Velocity statistics further behind the advancing edge}\label{vel_stat_further_behind}


\begin{figure}[h]
    \centering
    \includegraphics[width=1\linewidth]{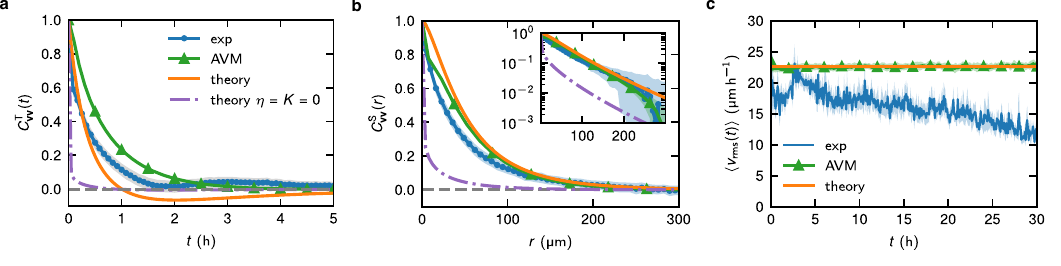}
    \caption{Velocity statistics in the combined experimental regions further behind the advancing edge, as defined and in Subsec. \ref{subsec:zoning}.  (\textbf{a}) The temporal autocorrelation function. (\textbf{b}) The spatial velocity correlation. (\textbf{c}) The root mean square velocity.}
    \label{fig:deeper_in_bulk_statisticsl}
\end{figure}

\subsection{Temporal autocorrelation of velocity field in expanding AVM}


\begin{figure}[h]
    \centering
    \includegraphics[width=0.667\linewidth]{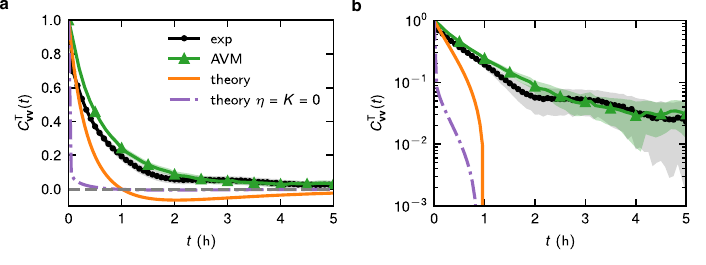}
    \caption{The temporal autocorrelation function of the velocity field  in the expanding AVM. (\textbf{a}) Linear scale (\textbf{b}) Log-lin scale.}
    \label{fig:autocorrelation_expanding_AVM}
\end{figure}

\clearpage

\subsection{Kymographs}\label{kymographs}

\begin{figure}[h]
    \centering
    \includegraphics[width=1\linewidth]{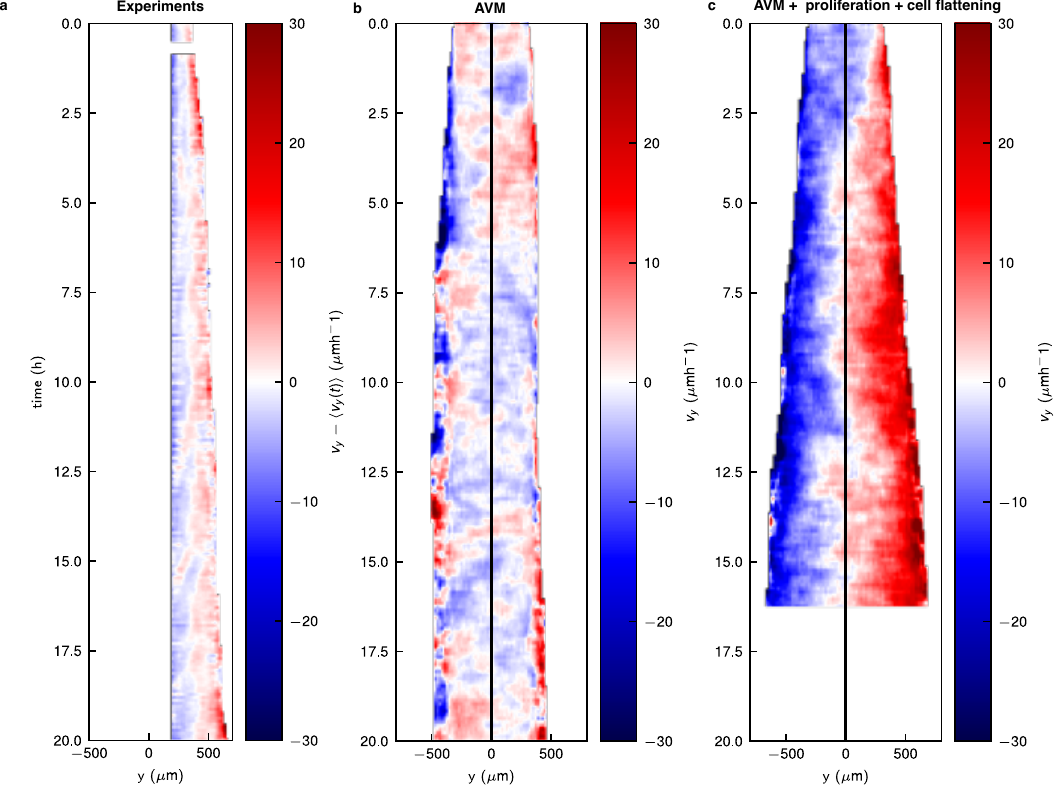} 
    \caption{Kymographs of $y$-component of the velocity averaged in strips running along $x$ with strip width $20$ $\mu$m.}
    \label{fig:kymographs}
\end{figure}
\clearpage
\subsection{Area, perimeter and shape index}\label{area_perimeter_shape_index}
\begin{figure}[h]
    \centering
    \includegraphics[width=0.7\linewidth]{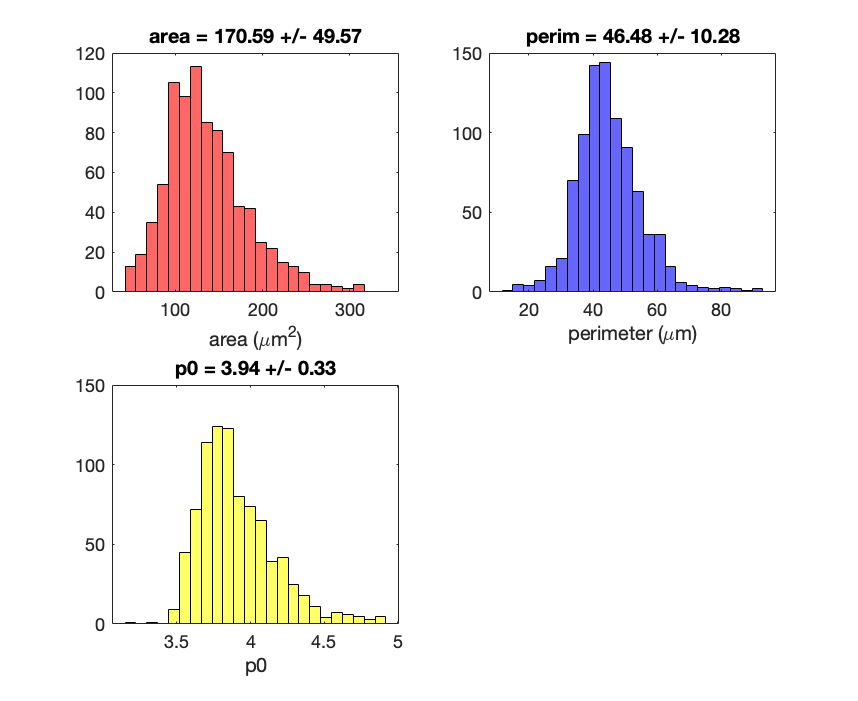}
    \caption{Distributions of area, perimeter, and shape index of frame 350 in experiment 07.}
    \label{fig:area_perimeter_shape_index_hist}
\end{figure}

\begin{figure}[h]
    \centering
    \includegraphics[width=1\linewidth]{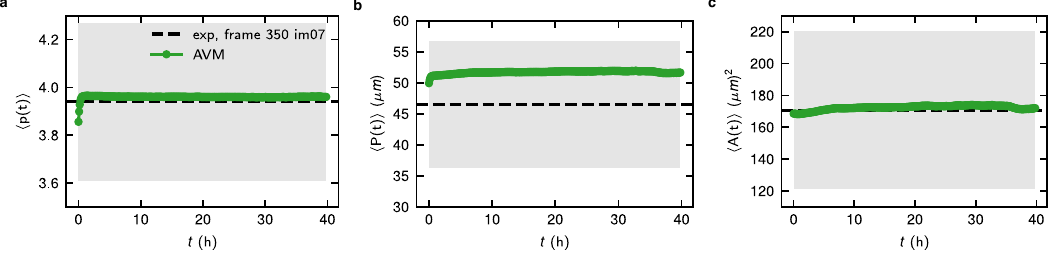}
    \caption{Evolution of area (\textbf{a}), perimeter (\textbf{b}), and shape index (\textbf{c}) of non-expanding AVM compared to  frame 350 in experiment 07. Shading of experiments is the standard deviation of experimental histograms and the shading of the AVM is the standard error of the mean. }
    \label{fig:area_perimeter_shape_index_evolution}
\end{figure}

\clearpage
\section{Boundary}

\subsection{Polymer simulations starting with initial roughness}
\begin{figure}[h]
    \centering
    \includegraphics[width=0.9\linewidth]{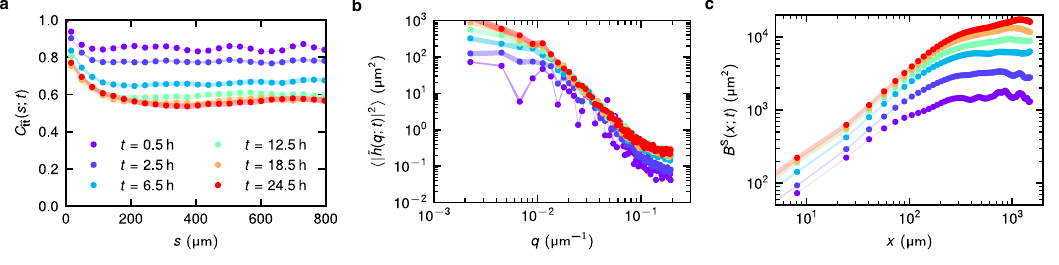}
    \caption{Tangent-tangent correlation (\textbf{a}), spatial Fourier transform (\textbf{b}), and relative roughness correlation (\textbf{c}) of polymer simulations that start with the inital boundary shape of experiment sequence 03, showing that initial roughness can indeed explain the very early time discrepancy with experiments.}
    \label{fig:polymer_sims_initial_roughness}
\end{figure}

\subsection{Early time boundary persistence}
\begin{figure}[h]
    \centering
    \includegraphics[width=0.6\linewidth]{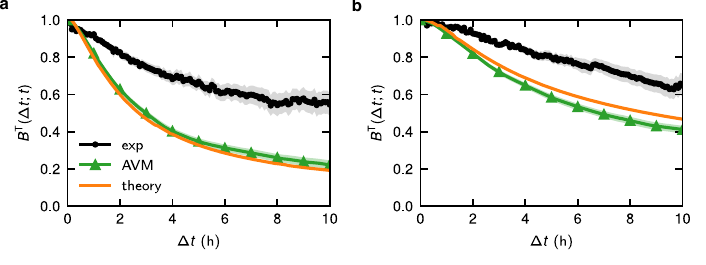}
    \caption{The autocorrelation function of the height at different times $t$. (\textbf{a}) $t=\SI{1}{\hour}$ (\textbf{b}) $t=\SI{6}{\hour}$.}
    \label{fig:early_time_boundary_persistence}
\end{figure}

\subsection{Finger formation in expanding AVM}
\begin{figure}[h]
    \centering
    \includegraphics[width=0.9\linewidth]{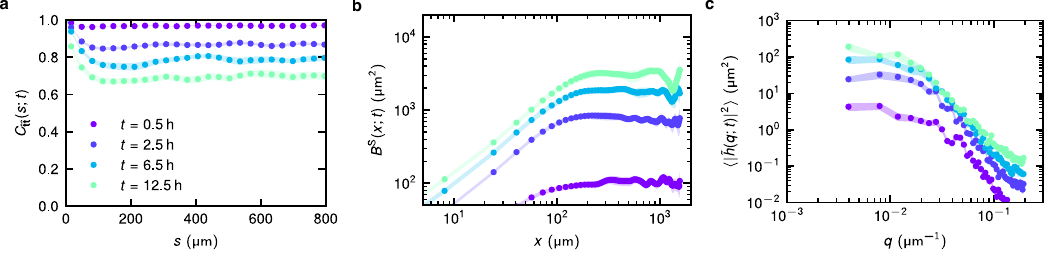}
    \caption{Tangent-tangent correlation (\textbf{a}), relative roughness correlation (\textbf{b}) and spatial Fourier transform (\textbf{c}) of the boundary of the expanding AVM simulations.}
    \label{fig:tangent_roughness_spectrum_expanding_AVM}
\end{figure}

\clearpage
\subsection{2d polymer velocity correlations}\label{fig:2d_polymer_vcors}
\begin{figure}[h]
    \centering
    \includegraphics[width=0.667\linewidth]{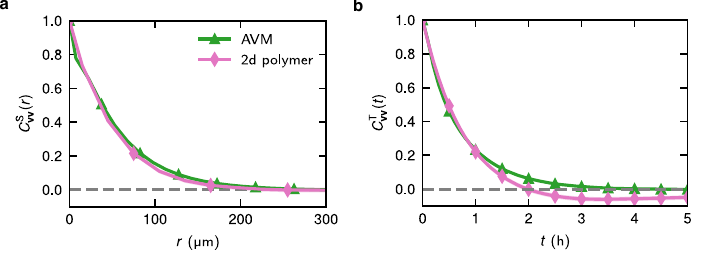}
    \caption{Real-space velocity correlations  of the noise-generating chain for the 2d polymer simulations and the velocity correlations in the AVM. (\textbf{a}) Spatial velocity correlation function. (\textbf{b}) Temporal autocorrelation function.  }
    \label{fig:enter-label}
\end{figure}

\part{Supporting videos}
\centering
\begin{enumerate}[label=\textbf{S\arabic*},ref=S\arabic*]
    \item  \label{mov:experiment_inverted_grayscale}
\path{experiment_inverted_compressed.mp4}. Experimental image sequence 01 with inverted greyscale over 30 \si{h}.
    \item  \label{mov:experiment_analyzed}
\path{experiment_analyzed_compressed.mp4}. Experimental image sequence 01, with extracted boundary (red line) and PIV field (black arrows) over 30 \si{h}.
    \item  \label{mov:non_expanding_AVM_full} \path{non_expanding_AVM.mp4} Movie of a typical non-expanding AVM simulation showing the full sheet over 30 \si{h}. The black arrows show the velocity vectors of the cells.
    \item  \label{mov:non_expanding_AVM_zoom} \path{non_expanding_AVM_zoomed.mp4} Movie of a typical non-expanding AVM simulation over 30 \si{h} showing a zoomed-in section of the sheet near the boundary. The black arrows show the velocity vectors of the cells.
    \item  \label{mov:expanding_AVM} \path{expanding_AVM.mp4} Movie of a typical expanding AVM simulation over  12 \si{h}. The black arrows show the velocity vectors of the cells.
    \item  \label{mov:polymer} \path{polymer.mp4} Example of a typical polymer simulation over 30 \si{h}. Due to the absence of repulsion between the beads in the polymer simulations (to keep the equivalence with the height equation) in combination with the strong driving, the polymer sometimes exhibits a loop.
    \item \label{mov:heq} \path{heq.mp4} Example of a typical simulation of the height equation over 30 \si{h}.
\end{enumerate}






\begin{justify}
\part{Theory}
\setcounter{section}{0}
\section{Velocity field of the interior}\label{vcor}


\subsection{Derivation of bulk velocity field}\label{2d_velocity_field_derivation}

The velocity field can be derived by modelling the cell sheet as an active (visco)elastic solid.
We follow the derivation \cite{henkesDenseActiveMatter2020} and add pair dissipation.
The equation of motion for the displacement field $\bm{u}$ within the cell sheet reads
 \begin{align}
     \zeta \bm{\dot{u}}(\bm{r},t) = \zeta v_0 \bm{\hat{n}}(\bm{r},t) + \nabla \cdot \bt{\sigma},
 \end{align}
 where $\zeta$ is a friction coefficient, which we write without the $c$ subscript of the main text for ease of notation, $\zeta v_0 \hat{\bm{n}}$ is a self-propulsion force field with $v_0$ a self-propulsion velocity.


 For an isotropic material, we associate a bulk modulus $B$ to uniform expansion and a shear modulus $\mu$ to shear deformation, hence we write the elastic part of the stress tensor as 

 \begin{align}
     \sigma_{\alpha \beta}^E = B \delta_{\alpha \beta} \varepsilon_{\gamma \gamma} + 2 \mu \left(\varepsilon_{\alpha \beta} - \frac{1}{d} \delta_{\alpha \beta}\varepsilon_{\gamma \gamma} \right),
\end{align}
where the (linear) strain tensor is given by $\varepsilon_{\alpha \beta} = \frac{1}{2} \left( \partial_{\beta}u_{\alpha}+\partial_{\alpha}u_{\beta}  \right)$.

Similarly, we write a viscous contribution to the stress tensor with bulk viscosity $K$ and shear viscosity $\eta$, 

\begin{align}
     \sigma_{\alpha \beta}^V = K \delta_{\alpha \beta} \dot{\gamma}_{\gamma \gamma} + 2 \eta \left(\dot{\gamma}_{\alpha \beta} - \frac{1}{d} \delta_{\alpha \beta}\dot{\gamma}_{\gamma \gamma} \right),
\end{align}
with the strain rate tensor $\dot{\gamma}_{\alpha\beta} = \frac{1}{2} \left( \partial_{\beta}\dot{u}_{\alpha}+\partial_{\alpha}\dot{u}_{\beta}  \right)$. Adding $ \nabla \cdot \bt{\sigma} = \nabla \cdot \bt{\sigma}^E + \nabla \cdot \bt{\sigma}^V $, we obtain the equation of motion for out viscoelastic material at the linear level, 

\begin{equation}
    \zeta \dot{\bm{u}}(\bm{r}, t) = \zeta v_0 \hat{\bm{n}}(\bm{r},t) + B \nabla (\nabla \cdot \bm{u}(\bm{r},t))  + \mu \nabla^2 \bm{u}(\bm{r},t) + K \nabla (\nabla \cdot \dot{\bm{u}}(\bm{r},t))  + \eta \nabla^2 \dot{\bm{u}}(\bm{r},t).
    \label{eq:interior_dyn}
\end{equation}

We introduce the Fourier transform
\begin{subequations}
\begin{align}
\bm{U}(\bm{q}, \omega) &= \int \dd t \, e^{\ii\omega t} \tilde{\bm{u}}(\bm{q}, t) =  \int \dd \bm{r} \int \dd t \, e^{\ii\bm{q}\cdot\bm{r} + \ii\omega t} \, \bm{u}(\bm{r}, t),\\
\bm{u}(\bm{r}, t) &= \frac{1}{(2\pi)^2} \int \dd \bm{q} \, e^{-\ii \bm{q} \cdot \bm{r}} \, \tilde{\bm{u}}(\bm{q}, t) = \frac{1}{(2\pi)^3} \int \dd \bm{q} \, e^{-\ii \bm{q} \cdot \bm{r}} \int \dd \omega \, e^{-\ii \bm{q}\cdot\bm{r} - \ii \omega t} \bm{U}(\bm{q}, \omega),
\end{align}
\label{Fourier transform conventions}%
\end{subequations}
and express Eq.\ \eqref{eq:interior_dyn} in the spatiotemporal Fourier space
\begin{align}
    -\ii\omega \zeta \bm{U}(\bm{q}, \omega) = \zeta v_0 \hat{\bm{N}}(\bm{q}, \omega) -(B-\ii \omega K) \left[\bm{U}(\bm{q}, \omega)\cdot \bm{q}\right]\bm{q} -(\mu - \ii \omega \eta) \left[\bm{q} \cdot \bm{q} \right]\bm{U}(\bm{q}, \omega).
    \label{eq:interior_dyn_fourier}
\end{align}
We introduce $\bm{U}_{\parallel} = (\bm{U}\cdot\hat{\bm{q}})\hat{\bm{q}}$ and $\bm{U}_{\perp} = \bm{U} - \bm{U}_{\parallel}$ the longitudinal and transverse components of $\bm{U}$, where $\hat{\bm{q}} = \bm{q}/q$ and $q = |\bm{q}|$, then rewrite Eq.\ \eqref{eq:interior_dyn_fourier}
\begin{align}
    -\ii\omega \zeta \begin{bmatrix}
           \bm{U}_{\parallel} \\
           \bm{U}_{\perp}
         \end{bmatrix}=\zeta v_0 \begin{bmatrix}
           \hat{\bm{N}}_{\parallel} \\
           \hat{\bm{N}}_{\perp}
         \end{bmatrix} -\begin{bmatrix}
           (B+\mu-\ii \omega (K+\eta)) q^2 & 0\\
           0 & (\mu - \ii \omega \eta)q^2
           \end{bmatrix}\begin{bmatrix}
           \bm{U}_{\parallel} \\
           \bm{U}_{\perp} 
         \end{bmatrix}.
\end{align}
This system of equations is decoupled for longitudinal and transverse components. We thus write
\begin{subequations}
\begin{align}
    \bm{U}_{\parallel}(\bm{q}, \omega) &= \frac{\zeta v_0 \bm{N}_{\parallel}(\bm{q}, \omega)}{\left[B+\mu \right]q^2-\left[K+\eta \right]\ii\omega q^2-\ii \omega \zeta},\\
    \label{transverse_mode}
    \bm{U}_{\perp}(\bm{q}, \omega) &= \frac{\zeta v_0 \bm{N}_{\perp}(\bm{q}, \omega)}{\mu q^2-\ii\eta \omega q^2-\ii \omega \zeta}.
\end{align}
\end{subequations}

\subsection{Temporal autocorrelation function}
\label{sec:cvvt}

We derive the temporal autocorrelation function. We first focus on the longitudinal component, and will infer the transverse component by identification while setting $B=0$ and $K = 0$.

We define the longitudinal velocity temporal autocorrelation function
\begin{equation}
    \la \bm{v}_{\parallel}(\bm{r},t) \cdot \bm{v}_{\parallel}(\bm{r},t') \ra = \frac{1}{(2 \pi)^6} \iiiint \dd\bm{q} \, \dd\bm{q'} \, \dd\omega \, \dd\omega' \, (-\omega \omega') \la \bm{U}_{\parallel}(\bm{q},\omega) \cdot \bm{U}_{\parallel}(\bm{q'},\omega') \ra e^{-\ii(\bm{q}+\bm{q'})\cdot\boldsymbol{r}}e^{-\ii\omega t} e^{-\ii\omega't'},
\end{equation}
and note that \cite{henkesDenseActiveMatter2020}
\begin{align}
    \la \hat{\bm{N}}_{\parallel}(\bm{q},\omega) \hat{\bm{N}}_{\parallel}(\bm{q'},\omega') \ra = \frac{1}{2} \la \hat{\bm{N}}(\bm{q},\omega) \hat{\bm{N}}(\bm{q'},\omega') \ra = (2 \pi)^3 a^2  \frac{\tau}{1+\omega^2 \tau^2} \delta(\bm{q}+\bm{q'})\delta(\omega+\omega'),
\end{align}
where $a$ is a coarse-graining length scale. We therefore obtain
\begin{align}
    \la \bm{v}_{\parallel}(\bm{r},t) \cdot \bm{v}_{\parallel}(\bm{r},t') \ra = \frac{a^2 \zeta^2 v_0^2 \tau}{(2 \pi)^3} \iint \dd\bm{q} \, \dd\omega  \, F(\bm{q},\omega) \,  e^{-\ii \omega (t-t')},
\label{eq:autovt}
\end{align}
with 
\begin{equation}
\begin{aligned}
    F(\bm{q},\omega) &= \frac{\omega^2}{\left[B+\mu \right]^2q^4+\left(\left[K+\eta \right]q^2+  \zeta\right)^2 \omega^2} \frac{1}{1+\tau^2 \omega^2} = \frac{\omega^2}{\frac{\zeta^2}{\tau^2}\Big[\frac{(B + \mu)^2 \tau^2}{\zeta^2}q^4 + \left(\frac{K + \eta}{\zeta}q^2 + 1\right) \tau^2 \omega^2\Big]\Big[1 + \tau^2 \omega^2\Big]}\\
    &= \frac{\tau^2\omega^2/\zeta^2}{\Big[(\xi_{\parallel,p}^2 - \xi_{\parallel,d}^2)^2 q^4 + (\xi_{\parallel,d}^2 q^2 + 1)^2 \tau^2 \omega^2\Big]\Big[1 + \tau^2\omega^2\Big]}\\
    &= \frac{1}{\zeta^2\Big[(\xi_{\parallel,p}^2 - \xi_{\parallel,d}^2)^2 q^4 - (\xi_{\parallel,d}^2 q^2 + 1)^2 \tau^2 \omega^2\Big]}\left[\frac{1}{1 + \frac{(\xi_{\parallel,d}^2 q^2 + 1)^2 \tau^2}{(\xi_{\parallel,p}^2 - \xi_{\parallel,d}^2)^2 q^4}\omega^2} - \frac{1}{1 + \tau^2 \omega^2}\right],
\end{aligned}
\label{eq:vparvparfqw}
\end{equation}
where we have introduced the following longitudinal length scales,
\begin{subequations}
\begin{align}
\xi_{\parallel,p} &= \sqrt{\frac{(B + \mu)\tau + K + \eta}{\zeta}},\\
\xi_{\parallel,d} &= \sqrt{\frac{K + \eta}{\zeta}}.
\end{align}
\label{eq:longitudinal_length_scales}%
\end{subequations}
We use the following relations between the Lorentzian and the exponential via the Fourier transform
\begin{subequations}
\begin{align}
\label{eq:ftexp}
\int_{-\infty}^{\infty} \dd t \, e^{\ii \omega t} \, e^{-|t|/\tau} &= \frac{2\tau}{1 + \tau^2 \omega^2},\\
\frac{1}{2\pi} \int_{-\infty}^{\infty} \dd \omega \, e^{-\ii \omega t} \frac{1}{1 + \tau^2 \omega^2} &= \frac{1}{2\tau} e^{-|t|/\tau},
\end{align}
\end{subequations}
therefore we take the inverse Fourier transform in time in Eq.\ \eqref{eq:autovt} as follows
\begin{equation}
\begin{aligned}
\la \bm{v}_{\parallel}(\bm{r},t) \cdot \bm{v}_{\parallel}(\bm{r},t') \ra &= \frac{a^2 v_0^2}{2 (2\pi)^2} \int \dd \bm{q} \, \frac{(\xi_{\parallel,p}^2 - \xi_{\parallel,d}^2) q^2 \, e^{-\frac{(\xi_{\parallel,p}^2 - \xi_{\parallel,d}^2) q^2}{(\xi_{\parallel,d}^2 q^2 + 1)\tau}|t - t^{\prime}|} - (\xi_{\parallel,d}^2 q^2 + 1) \, e^{-|t - t^{\prime}|/\tau}}{\Big[(\xi_{\parallel,p}^2 - \xi_{\parallel,d}^2)^2 q^4 - (\xi_{\parallel,d}^2 q^2 + 1)^2\Big]\Big[\xi_{\parallel,d}^2 q^2 + 1\Big]}\\
&= \frac{a^2 v_0^2}{4\pi} \int \dd q \, q \, \frac{(\xi_{\parallel,p}^2 - \xi_{\parallel,d}^2) q^2 \, e^{-\frac{(\xi_{\parallel,p}^2 - \xi_{\parallel,d}^2) q^2}{(\xi_{\parallel,d}^2 q^2 + 1)\tau}|t - t^{\prime}|} - (\xi_{\parallel,d}^2 q^2 + 1) \, e^{-|t - t^{\prime}|/\tau}}{\Big[(\xi_{\parallel,p}^2 - \xi_{\parallel,d}^2)^2 q^4 - (\xi_{\parallel,d}^2 q^2 + 1)^2\Big]\Big[\xi_{\parallel,d}^2 q^2 + 1\Big]}\\
&= \frac{a^2 v_0^2}{4 \pi \tau}  \int_0^{\infty} dq \, q\, \frac{\zeta^2}{\zeta_{q,\parallel}^2}\frac{\frac{ (B+\mu)q^2 }{\zeta_{q,\parallel}} e^{-(B+\mu)q^2|t-t'|/\zeta_{q,\parallel}} -\frac{1}{\tau}e^{-|t-t'|/\tau}}{\left[ (B+\mu)q^2/\zeta_{q,\parallel}\right]^2 - 1/\tau^2},
\end{aligned}
\label{eq:vparallelautocorr}%
\end{equation}
where we have introduced
\begin{equation}
    \zeta_{q,\parallel} = \zeta + (K + \eta) q^2.
    \label{eq:zetaq}
\end{equation}
For $\la \bm{v}_{\perp}(\bm{r},t) \cdot \bm{v}_{\perp}(\bm{r},t') \ra$ we set $B=0$ and $K=0$ and we then combine the results to find the full temporal autocorrelation function: $\la \bm{v}(\bm{r},t) \cdot \bm{v}(\bm{r},t') \ra =  \la \bm{v}_{\parallel}(\bm{r},t) \cdot \bm{v}_{\parallel}(\bm{r},t') \ra + \la \bm{v}_{\perp}(\bm{r},t) \cdot \bm{v}_{\perp}(\bm{r},t')\ra $.

\subsection{Spatial correlation function}
\label{sec:cvvr}

Using the notations of Eq.\ \eqref{eq:autovt}, we write the velocity spatial correlation function
\begin{equation}
    \la \bm{v}_{\parallel}(\bm{r},t) \cdot \bm{v}_{\parallel}(\bm{r}',t) \ra = \frac{a^2 \zeta^2 v_0^2 \tau}{(2 \pi)^3} \iint \dd\bm{q} \, \dd\omega  \, F(\bm{q},\omega) \,  e^{-\ii \bm{q} (\bm{r}-\bm{r}')},
\end{equation}
and take the inverse Fourier transform in time similarly to Eq.\ \eqref{eq:vparallelautocorr} with $t = t^{\prime}$
\begin{equation}
\begin{aligned}
    \la \bm{v}_{\parallel}(\bm{r},t) \cdot \bm{v}_{\parallel}(\bm{r}',t) \ra &= \frac{a^2 v_0^2}{2 (2 \pi)^2} \int \dd \bm{q} \, e^{-\ii \bm{q} (\bm{r}-\bm{r}')} \, \frac{(\xi_{\parallel,p}^2 - \xi_{\parallel,d}^2) q^2 - (\xi_{\parallel,d}^2 q^2 + 1)}{\Big[(\xi_{\parallel,p}^2 - \xi_{\parallel,d}^2)^2 q^4 - (\xi_{\parallel,d}^2 q^2 + 1)^2\Big]\Big[\xi_{\parallel,d}^2 q^2 + 1\Big]}\\
    &= \frac{a^2 v_0^2}{2 (2 \pi)^2} \int \dd \bm{q} \, e^{-\ii \bm{q} (\bm{r}-\bm{r}')} \, \frac{1}{\Big[\xi_{\parallel,p}^2 q^2 + 1\Big]\Big[\xi_{\parallel,d}^2 q^2 + 1\Big]}\\
    &= \frac{a^2 v_0^2}{4 \pi} \int \dd q \, q \, J_0(q |\bm{r} - \bm{r}'|) \, \frac{1}{\Big[\xi_{\parallel,p}^2 q^2 + 1\Big]\Big[\xi_{\parallel,d}^2 q^2 + 1\Big]},
\end{aligned}
\end{equation}
where $J_0$ is the 0-th Bessel function of the first kind, and finally take the inverse Fourier transform in space
\begin{equation}
    \la \bm{v}_{\parallel}(\bm{r},t) \cdot \bm{v}_{\parallel}(\bm{r}',t) \ra = \frac{a^2 v_0^2}{4 \pi (\xi_{\parallel,p}^2 - \xi_{\parallel,d}^2)} [K_0(r/\xi_{\parallel,p}) - K_0(r/\xi_{\parallel,d})] \underset{\bm{r} = \bm{r}'}{=} \frac{a^2 v_0^2}{4 \pi (\xi_{\parallel,p}^2 - \xi_{\parallel,d}^2)} \log(\xi_{\parallel,p}/\xi_{\parallel,d}),
\end{equation}
where $K_0$ is the 0-th modified Bessel function of the second kind and $r = |\bm{r} - \bm{r}'|$.

Similarly to Eq.\ \eqref{eq:longitudinal_length_scales}, we introduce the transverse length scales
\begin{subequations}
\begin{align}
\xi_{\perp,p} &= \sqrt{\frac{\mu\tau + \eta}{\zeta}},\\
\xi_{\perp,d} &= \sqrt{\frac{\eta}{\zeta}}.
\end{align}
\label{eq:translengths}%
\end{subequations}
and write the full velocity spatial correlation function
\begin{equation}
\begin{aligned}
    \la \bm{v}(\bm{r},t) \cdot \bm{v}(\bm{r}',t) \ra &= \la \bm{v}_{\parallel}(\bm{r},t) \cdot \bm{v}_{\parallel}(\bm{r}',t) \ra + \la \bm{v}_{\perp}(\bm{r},t) \cdot \bm{v}_{\perp}(\bm{r}',t) \ra\\
    &= \frac{a^2 v_0^2}{4 \pi} \left[\frac{K_0(r/\xi_{\parallel,p}) - K_0(r/\xi_{\parallel,d})}{\xi_{\parallel,p}^2 - \xi_{\parallel,d}^2} + \frac{K_0(r/\xi_{\perp,p}) - K_0(r/\xi_{\perp,d})}{\xi_{\perp,p}^2 - \xi_{\perp,d}^2}\right].
\end{aligned}
\end{equation}

It is noteworthy that in the absence of pair dissipation $K = \eta = 0$ and thus $\xi_{\parallel,d} = \xi_{\perp,d} = 0$ which leads to
\begin{equation}
    \la \bm{v}(\bm{r},t) \cdot \bm{v}(\bm{r}',t) \ra = \frac{a^2 v_0^2}{4 \pi} \left[\frac{K_0(r/\xi_{\parallel,p})}{\xi_{\parallel,p}^2} + \frac{K_0(r/\xi_{\perp,p})}{\xi_{\perp,p}^2}\right],
\end{equation}
where $\lim_{r\to0} K_0(r/\xi) = \infty$ for any $\xi > 0$.

\subsection{Mapping AVM interior parameters to bulk velocity field}\label{AVM_interior_mapping}
\begin{table}[b!]
    \centering
    \begin{tabular}{c|c}
        Parameter & Units \\
        \hline
         $B/\zeta_c$& $L^2 T^{-1}$ \\
         $\mu/\zeta_c$& $L^2 T^{-1}$\\
         $K/\zeta_c$& $L^2$\\
         $\eta/\zeta_c$& $L^2$\\
         $\tau$& $T$ \\
         $v_0$& $L T^{-1}$ \\
    \end{tabular}
    \caption{Parameters and units of the 2d continuum active viscoelastic model for the bulk velocity field of the cell sheet.}
    \label{tab:cont_units}
\end{table}
Here we explain how the parameters of the AVM are mapped to the continuum viscoelastic model parameters. We do this based on dimensional analysis and a reference length scale $\sqrt{A_0}$. For clarity, we write the friction coefficient of the continuum model ($\zeta$ in the previous subsection) with a subscript: $\zeta_c$ and, like in the main text, take the friction coefficient of the AVM as $\zeta$.
We deduce from Eq.\ \eqref{eq:interior_dyn} the dimensions of the continuum parameters.
We take $\tau$ and $v_0$ the same in AVM and the continuum model.
Based on the dimensions in table \ref{tab:cont_units} , we take $\mu/\zeta_c = A_0 \Gamma/\zeta_c $ and $\eta/\zeta_c = A_0\zeta_{\mathrm{pair}}/\zeta $. Furthermore, we assume $B=2\mu$ and $K=2\eta$.

\clearpage
\newpage
\section{Derivation of the height equation}\label{heq_derivation}

This document derives the height equation theory for the cell sheet boundary. The motivation for studying the height equation is on the one hand to validate the hypothesis that correlated cell movements driving the contractile boundary leads to finger formation and then to lay out the relation between the different time scales (\textit{e.g.} the short persistence time of cell driving and the long persistence time of fingers) and length scales. On the other hand, this study aims to provide quantitative comparisons between experiment, AVM, and theory.

The latter goal constrains the setup of the height equation.
In the experiments and AVM simulations, we only have access to a piece of the boundary, namely, a piece of horizontal length $L_x$. The idea is to treat this piece of the boundary as a worm-like chain of horizontal length $L_x$ with periodic boundary conditions.

For the sake of clarity, we first quote the equation of motion in time $t$ for the single-valued vertical height $h$ of the boundary as a function of the horizontal direction $x$,
\begin{equation}
    \zeta_h \partial_t h(x,t) = \lambda \partial_x^2h(x,t) - \kappa \partial_x^4 h(x,t) + \eta_h \partial_x^2 \partial_t h(x,t)  + \zeta_h v^{f}(x,t),
\label{height equation, pair-dissipation}
\end{equation}
referred to below as the height equation, and then explain the meaning of each term.
We assume overdamped dynamics for the boundary with friction coefficient $\zeta_h$. The line tension $\lambda$ and bending stiffness $\kappa$ contribute the first two terms on the right-hand side and these terms will be derived from a worm-like chain in section \ref{WLC}. Motivated by the bulk dynamics and the AVM model, we also add pair-dissipation (through viscosity $\eta_h$) to the height equation. The boundary is driven by the stochastic velocity field $v^{f}(x,t)$ from the bulk of the cell sheet. The characterisation of this stochastic velocity field by its correlation functions can be found in section \ref{vcor}.

\subsection{Worm-like chain contribution}\label{WLC}

We consider a worm-like chain model of a polymer, parametrise the position along the polymer with the arc length $s$, and write its free energy \cite{saarloosSoftMatterConcepts2024} as 
\begin{equation}
    E_{\text{uns}} = \int_0^L \dd s \, \frac{1}{2} \kappa\left(\partial_s^2 \bm{r} \right)^2,
\end{equation}
where $L$ is the total length of the polymer and $\kappa$ its bending stiffness.


We stretch the polymer to a horizontal length $L_x$. This global constraint \cite{byronMathematicsClassicalQuantum2012} can be written as 
\begin{equation}
    \int_0 ^L  \dd s \, \partial_s \bm{r}\cdot \buv{x} = L_x,
\end{equation}
and can be enforced by adding it to the free energy with a Lagrange multiplier $\lambda$
\begin{equation}
     E =  \int_0^L \dd s \, \frac{1}{2} \kappa\left(\partial_s^2 \bm{r} \right)^2 - \lambda\left(\int_0 ^L \dd s \,  \partial_s \bm{r}\cdot \buv{x} -L_x\right).
     \label{strectched worm-like chain energy}
\end{equation}

We are interested in the fluctuations with respect to an initial horizontal configuration, i.e.\ $\bm{r}(s,t=0)=x(s)\buv{x}$. We therefore decompose $\bm{r}(s)$ into a parallel (along $x$-axis) and a perpendicular (along $y$-axis) component \cite{liverpoolAnomalousFluctuationsActive2003}
\begin{equation}
    \bm{r}(s) = (s- r_{\parallel}(s)) \buv{x} + r_{\perp}(s) \buv{y},
\end{equation}
which also yields
\begin{subequations}
\begin{align}
    \partial_s\bm{r}(s) &= (1- \partial_s r_{\parallel}) \buv{x} + \partial_s r_{\perp} \buv{y},\\
    \partial_s^2\bm{r}(s) &= - \partial_s^2 r_{\parallel} \buv{x} + \partial_s^2 r_{\perp} \buv{y}.
\end{align}
\end{subequations}

Using this decomposition, the free energy Eq.\ \eqref{strectched worm-like chain energy} takes the form
\begin{equation}
       E = \int_0 ^L \dd s \, \left(  \frac{\kappa}{2}\left[(\partial_s^2 r_{\parallel})^2 +  (\partial_s^2 r_{\perp})^2\right] - \lambda(1- \partial_s r_{\parallel}) \right) + \lambda L_x,
\end{equation}
which is the free energy of the polymer only if $s$ is the arc length. This requirement constrains the fluctuations as the tangent vector has to be of unit norm:
\begin{equation}
    (1-\partial_s r_\parallel)^2 + (\partial_s r_\perp)^2 = 1.
    \label{eq:local_fluc_constraint}
\end{equation}

\subsection{Limit of small fluctuations}

Note that Eq.\ \eqref{eq:local_fluc_constraint} is a \textit{local} constraint and generally allows expressing one of the degrees of freedom in terms of the others, reducing the total number of degrees of freedom \cite{byronMathematicsClassicalQuantum2012}. In the limit of small fluctuations, this is straightforward to do by solving the constraint equation for $\partial_s r_\parallel$,
\begin{equation}
    \partial_s r_\parallel = 1 - \sqrt{1 - (\partial_s r_\perp)^2}
    = \frac{1}{2} (\partial_s r_\perp)^2 + \frac{1}{8}(\partial_s r_\perp)^4 + \mathcal{O}((\partial_s r_\perp)^6),
\end{equation}
and then approximating $\partial_s r_\parallel \approx \frac{1}{2} (\partial_s r_\perp)^2$.

In the limit of small fluctuation, we thus obtain, dropping the constant term $\lambda L_x$, the following expression for the free energy
\begin{equation}
    E = \int_0^L \dd s \, \left( \frac{\kappa}{2} (\partial_s^2 r_{\perp})^2 + \frac{\lambda}{2}(\partial_s r_{\perp})^2 - \lambda \right)
    \equiv \int_0^L \dd s \, f(s, \partial_s r_{\perp}, \partial_s^2 r_{\perp}).
\end{equation}
Passive forces are then given by the variation of this free energy
\begin{equation}
    \frac{\delta E}{\delta r_{\perp}} = \frac{\partial f}{\partial r_{\perp}} - \frac{\dd}{\dd s}\left(\frac{\partial f}{\partial \partial_s r_{\perp}} \right) + \frac{\dd^2}{\dd s^2} \left(\frac{\partial f}{\partial \partial_s^2 r_{\perp}} \right)
    = -\lambda \partial_s^2 r_{\perp} + \kappa \partial_s^4 r_{\perp}.
\end{equation}
Note that we used here that $\lambda $ is constant. 
In the absence of overhangs, we can write $s(x)$, and in the limit of small fluctuations,  $\partial_s x = 1 - \partial_s r_{\parallel}$, so that higher order derivatives with respect to $s$ can be approximated as derivatives with respect to $x$. Thus, after identifying $h(x,t) \equiv r_{\perp}(x,t)$, we finally obtain
\begin{equation}
    \frac{\delta E}{ \delta h} = -\lambda \partial_x^2 h(x,t) + \kappa \partial_x^4 h(x,t).
    \label{energy_variation}
\end{equation}

\subsection{From 2d to 1d}

We consider that the bulk dynamics generate the driving fluctuating velocity field $v^f$ of the boundary.
In the case of a flat boundary, \textit{e.g.} oriented in the direction of $\hat{\bm{x}}$, we would expect these fluctuations to have the following spectrum
\begin{equation}
    \la V^f(q, \omega) V^f(q', \omega') \ra = \iiiint \dd x \, \dd t \, \dd x' \, \dd t' \, e^{\ii q x + \ii \omega t} \, e^{\ii q' x' + \ii \omega' t'} \, \la (\bm{v}(\bm{r}, t) \cdot \hat{\bm{y}})(\bm{v}(\bm{r} + x \hat{\bm{x}}, t')\cdot \hat{\bm{y}}) \ra.
\end{equation}
We introduce $\theta$ such that $\bm{q} = q \cos\theta \, \hat{\bm{x}} + q \sin\theta \, \hat{\bm{y}}$, then we write the above velocity correlations
\begin{equation}
\begin{aligned}
    &\la (\bm{v}(\bm{r}, t) \cdot \hat{\bm{y}})(\bm{v}(\bm{r} + x \hat{\bm{x}}, t')\cdot \hat{\bm{y}}) \ra\\
    &\qquad= \frac{1}{(2\pi)^4} \iint \dd \bm{q} \, \dd \bm{q}' \, e^{-\ii\bm{q}\cdot\bm{r}} e^{-\ii\bm{q}'\cdot(\bm{r} + x \hat{\bm{x}})} \, \left[\la \tilde{\bm{v}}_{\parallel}(\bm{q}, t) \cdot \tilde{\bm{v}}_{\parallel}(\bm{q}', t') \ra \cos^2 \theta + \la \tilde{\bm{v}}_{\perp}(\bm{q}, t) \cdot \tilde{\bm{v}}_{\perp}(\bm{q}', t') \ra \sin^2 \theta \right],
\end{aligned}
\label{eq:vyvy}
\end{equation}
where we have assumed $\la \tilde{\bm{v}}_{\parallel}(\bm{q}, t) \cdot \tilde{\bm{v}}_{\perp}(\bm{q}', t') \ra = 0$.
Given the derivations of sections \ref{sec:cvvr} and \ref{sec:cvvt}, we write the spectra as following
\begin{subequations}
\begin{align}
    \la \tilde{\bm{v}}_{\parallel}(\bm{q}, t) \cdot \tilde{\bm{v}}_{\parallel}(\bm{q}', t') \ra &= G_{\parallel}(q, |t - t'|) \, \delta(\bm{q} + \bm{q}'),\\
    \la \tilde{\bm{v}}_{\perp}(\bm{q}, t) \cdot \tilde{\bm{v}}_{\perp}(\bm{q}', t') \ra &= G_{\perp}(q, |t - t'|) \, \delta(\bm{q} + \bm{q}'),
\end{align}
\end{subequations}
such that we may write Eq.\ \eqref{eq:vyvy} as
\begin{equation}
\begin{aligned}
    &\la (\bm{v}(\bm{r}, t) \cdot \hat{\bm{y}})(\bm{v}(\bm{r} + x \hat{\bm{x}}, t')\cdot \hat{\bm{y}}) \ra = \frac{1}{(2\pi)^4} \iint \dd q \, q \, \dd \theta \, e^{-\ii q x \cos\theta} \left[G_{\parallel}(q, |t - t'|)\cos^2\theta + G_{\perp}(q, |t - t'|) \sin^2\theta \right]\\
    &\qquad= \frac{1}{2 (2\pi)^3} \int \dd q \, q \, 
    \left[G_{\parallel}(q, |t - t'|) \, \big(J_0(|q x|) + J_2(|qx|)\big) + G_{\perp}(q, |t - t'|) \, \big(J_0(|q x|) - J_2(|qx|)\big)\right],
\end{aligned}
\label{eq:vyvyfull}
\end{equation}
where we have used
\begin{equation}
    \int_0^{2\pi} \dd \theta e^{-\ii q x \cos\theta}\cos^2\theta = \frac{1}{2} 2\pi (J_0(|qx|) + J_2(|q x|)),
\end{equation}
with $J_2$ the 2\textsuperscript{nd} Bessel function of the first kind.
Crucially, Eq.\ \eqref{eq:vyvyfull} shows that
\begin{equation}
    \la (\bm{v}(\bm{r}, t) \cdot \hat{\bm{y}})(\bm{v}(\bm{r} + x \hat{\bm{x}}, t')\cdot \hat{\bm{y}}) \ra \neq \la \bm{v}_{\perp}(\bm{r}, t) \cdot \bm{v}_{\perp}(\bm{r} + x \hat{\bm{x}}, t')\ra = \frac{1}{(2\pi)^3} \int \dd q \ q \, G_{\perp}(q, |t - t'|) \,  J_0(|q x|),
\end{equation}
therefore we cannot directly use the fluctuations of the 2d bulk velocity field to inform 1d boundary fluctuations.

Therefore, as  a matter a simplicity, we will consider that the driving fluctuating velocity field $v^f$ is generated by the following chain model
\begin{equation}
     \zeta v^f (x,t)= \zeta \partial_t y(x,t) = \mu \partial_x^2 y(x,t) +\eta \partial_x^2 \partial_t  y(x,t)+  \zeta v_0 n_y (x,t).
\label{eom support chain pair, xt space}
\end{equation}
In this equation, $\zeta v_0n_y (x,t)$ is the $y$-component of the ABP active driving field in \cite{henkesDenseActiveMatter2020}. It is correlated in time, and uncorrelated in space
\begin{align}
    \la n_y (x,t )n_y (x',t') \ra = \frac{1}{2} a \, \delta(x-x') \, e^{-|t-t'|/\tau},
\label{eq:nyny}
\end{align}
with $a$ a coarse-graining length equivalent to the lattice spacing (see the supplementary of  \cite{henkesDenseActiveMatter2020} for more details).

We write Eq.\ \eqref{eom support chain pair, xt space} in $(q, \omega)$-space
\begin{equation}
    -\ii \zeta \omega Y(q, \omega) = -\mu q^2Y(q, \omega) + \ii \eta \omega q^2 Y(q, \omega)+\zeta v_0 N_y(q, \omega),
\end{equation}
which yields
\begin{equation}
     V^f(q,\omega)= -\ii \omega Y(q, \omega) = \frac{- \ii \omega \zeta v_0 N_y(q, \omega)}{\mu q^2 - \ii \eta \omega q^2 - \ii \zeta \omega}.
     \label{Vf(q,omega)}
\end{equation}

\subsection{Mapping AVM boundary to height equation + noise-generating chain}\label{mapping_AVM_boundary_to_heq_and_polymer}

Here, we explain the mapping of the  AVM boundary parameters to the height equation + noise-generating chain parameters. We do so by coarse-graining the boundary polymer of the AVM to the height equation. For simplicity, we only consider the forces between the boundary points and assume that the interaction with the interior can be modelled by the advection with the velocity field of the interior. 
We define the position vector of each boundary point by 
$\bm{r}_i$, where $i$ runs from 1 to $N$, the number of boundary points. The link/bond vectors are defined as $\bm{R}_i = \bm{r}_{i+1} - \bm{r}_{i}$, where $i$ runs from 1  to $N$.
\begin{figure}[h]
    \centering
    \includegraphics[width=0.4\linewidth]{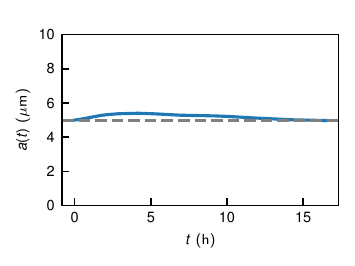}
    \caption{Evolution of the mean distance between boundary points in the AVM for matched simulations of the main text.}
    \label{fig:a_evolution}
\end{figure}

The linear spring potential with zero rest length of the AVM can be written as

\begin{equation}
    U_s = \frac{k_s}{2} \sum_{i=1}^{N} |\bm{R}_i|^2.
\end{equation}

The angular spring potential in the AVM is based on the angle $\theta_i$ between two link vectors and is given by 

\begin{align}
    U_{b, \mathrm{AVM}} =\frac{k_{b,\mathrm{AVM}}}{2}\sum_{i=0}^{N-2}\theta_i^2.
\end{align}

This bending potential does not map exactly to the fourth order bending term in the height equation. For that end, we have to instead take 

\begin{equation}
    U_b = \frac{k_b}{2} \sum_{i=1}^{N} \left( \bm{R}_{i+1} - \bm{R}_{i}\right)^2,
\end{equation}
which does approximate the bending potential of the AVM when the angle $\theta$ between link vectors is small (limit of small fluctuations) and $R_i\approx R_{i+1} \approx R\equiv a$ (constant tension):
\begin{equation}
    U_b  = \frac{k_b}{2} \sum_{i=1}^{N} \left[ |\bm{R}_{i+1}|^2 + |\bm{R}_{i}|^2 -2 |\bm{R}_{i+1}||\bm{R}_{i}|\cos(\theta_i) \right] 
     \approx \frac{k_b}{2} 2\sum_{i=0}^{N-2} \left[R^2 -R^2 \left(1-\frac{1}{2}\theta_i^2 \right)\right] = \frac{k_b}{2}R^2 \sum_{i=0}^{N-2}\theta_i^2.
\end{equation}
In figure \ref{fig:a_evolution}, we show that we can indeed approximate the bond lengths by a constant $a(t)\approx a(0)=5$ \si{\micro\meter} in the matched AVM simulations.

For periodic boundary conditions, one can then show (using the gradients of $U_s$ and $U_b$) that the equation of motion for each boundary point is given by
\begin{align}
        \zeta \dot{\bm{r}}_i = k_s \left(\bm{r}_{i+1}-2\bm{r}_{i} + \bm{r}_{i-1} \right) -k_b \left( \bm{r}_{i+2} -4   \bm{r}_{i+1} + 6  \bm{r}_{i} -4   
        \bm{r}_{i-1}+\bm{r}_{i-2} \right) +\zeta_{\mathrm{pair}} \left(\bm{r}_{i+1}-2\bm{r}_{i} + \bm{r}_{i-1} \right)
    + \bm{f}_i,
\end{align} 
where $\bm{f}_i = \zeta \bm{v}(\bm{r}_i)$ is the advection by the velocity field of the interior at position $\bm{r}_i$.

Dividing by the friction coefficient in the AVM and approximating the discrete differences by second and fourth order derivatives, we obtain the approximation

\begin{align}
        \dot{\bm{r}}(s) = \frac{k_s}{\zeta a^2} \partial_s^2 \bm{r}(s) - \frac{k_b}{\zeta a^4} \partial_s^4 \bm{r}(s) +  \frac{\zeta_{\mathrm{pair}}}{\zeta a^2} \partial_s^2 \partial_t\bm{r}(s)
    + \frac{\bm{f}(s)}{\zeta}.
\end{align} 
As explained in the derivation of the height equation, section \ref{heq_derivation}, we thus get in the limit of small fluctuations around an initially flat boundary:

\begin{align}
    \partial_t h(x,t) =  \frac{k_s}{\zeta a^2} \partial_x^2 h(x,t) - \frac{k_b}{\zeta a^4} \partial_x^4 h(x,t) +  \frac{\zeta_{\mathrm{pair}}}{\zeta a^2} \partial_x^2 \partial_t h(x,t)+  v^{f}(x,t),
\end{align}
from which we deduce the discrete-to-continuum mapping $ \frac{k_s}{\zeta a^2} = \frac{\lambda}{\zeta_h}$, $ \frac{k_b}{\zeta a^4} = \frac{\kappa}{\zeta_h} = \frac{k_{b,\mathrm{AVM}}}{\zeta a^2}$, and $\frac{\zeta_{\mathrm{pair}}}{\zeta a^2} = \frac{\eta_h}{\zeta_h}$. However, we found that the estimated line tension and bending stiffness on the boundary of the AVM should be taken roughly a factor of 25 smaller, to match with theory, likely stemming from the relaxation effects when new points are added to the AVM boundary, see \cite{bartonActiveVertexModel2017} for details.

We note that the spacing between boundary points $a(t)\approx a(0)=5$ \si{\micro\meter} in the AVM, is different from the coarse-graining length scale $\sqrt{A_0}$ used for the interior length scale in section \ref{AVM_interior_mapping}. This has implications for the mapping of the AVM parameters to the height equation noise-generating chain,
\begin{equation}
     v^f (x,t)=  \frac{\mu_h}{\zeta_h} \partial_x^2 y(x,t) + \frac{\eta_h}{\zeta_h} \partial_x^2 \partial_t  y(x,t)+   v_{0,h} n_y (x,t),
\label{eom support chain, parameters}
\end{equation}
as we will now discuss.

First of all, the driving strength $v_{0,h}$ is not equal to $v_0$ used in the AVM, this is because the driving of the boundary points by the cells of the interior is on a length scale $\sqrt{A_0}$ instead of the boundary length scale $a$. By discretizing Eq.\ \eqref{eom support chain, parameters} on the length scale $\sqrt{A_0}$ and then reparametrizing to the length scale $a$, one finds that $v_{0,h}= \frac{a}{\sqrt{A_0}}v_0$. Strictly speaking, this method also results in a viscosity parameter that is $\frac{a^2}{A_0}$ smaller, but we approximate this to be the same as the viscosity term of the height equation, and checked for the height spectrum that it made only a small difference at scales smaller than the cell size.  We use the equivalent $2d$ version of this support chain for the $2d$ polymer simulations mentioned in the main text, and find $\mu_h/\zeta_h=\SI{600}{\micro \meter^2\per\hour}$ by hand from fitting the real space spatial correlation of the 2d polymer noise generating chain velocity to the real space velocity correlations of the AVM (Supporting Figures, \ref{fig:2d_polymer_vcors}).


\subsection{Steady state velocity correlations}

We compute the steady state velocity correlation functions which enter the analysis of the height equation.
To that end we first compute the spectrum of the driving field $n_y$
\begin{equation}
    \la  N_y (q,\omega )N_y (q',\omega') \ra = \iiiint \dd x \, \dd t \, \dd x' \, \dd t' \, \la  n_y (x, t) n_y (x', t') \ra = 4  \pi^2 \tau a \frac{\delta(q+q')\delta(\omega+\omega')}{1+\tau^2 \omega^2},
    \label{n n, q omega space}
\end{equation}
where the second equality derives from Eqs.\ \eqref{eq:nyny} and \eqref{eq:ftexp}.

With the use of Eqs.\ \eqref{Vf(q,omega)} and  \eqref{n n, q omega space}, we readily obtain
\begin{equation}
\begin{aligned}
    \la V^f(q,\omega) V^f(q',\omega') \ra &= 
    \frac{4 \pi^2 \tau a \omega^2 \zeta^2 v_0^2}{\Big[\mu^2 q^4 +\left(\zeta + \eta q^2\right)^2 \omega^2\Big] \Big[1 + \tau^2 \omega^2 \Big]} \, \delta(q+q') \, \delta(\omega+\omega')\\
    &= 4\pi^2\tau a v_0^2 \zeta^2 \frac{\tau^2 \omega^2/\zeta^2}{\Big[(\xi_p^2 - \xi_d^2)^2 q^4 + (\xi_d^2q^2 + 1) \tau^2\omega^2\Big]\Big[1 + \tau^2\omega^2\Big]} \, \delta(q+q') \, \delta(\omega+\omega')
\label{eq: vv, q, omega, pair-dissipation}
\end{aligned}
\end{equation}
which has the same form as Eq.\ \eqref{eq:vparvparfqw} and where we have introduced the following length scales
\begin{subequations}
\begin{align}
    \xi_p &= \sqrt{\frac{\mu \tau + \eta}{\zeta}},\\
    \xi_d &= \frac{\eta}{\zeta},
\end{align}
\end{subequations}
identically to Eq.\ \eqref{eq:translengths}.
Similarly to Eq.\ \eqref{eq:vparallelautocorr}, we then compute the two-time velocity spatial spectrum
\begin{equation}
\begin{aligned}
    \la \tilde{v}^f(q,t) \tilde{v}^f(q',t')\ra &= \pi a v_0^2 \, \delta(q + q') \,  \frac{(\xi_{p}^2 - \xi_{d}^2) q^2 \, e^{-\frac{(\xi_{p}^2 - \xi_{d}^2) q^2}{(\xi_{d}^2 q^2 + 1)\tau}|t - t^{\prime}|} - (\xi_{d}^2 q^2 + 1) \, e^{-|t - t^{\prime}|/\tau}}{\Big[(\xi_{p}^2 - \xi_{d}^2)^2 q^4 - (\xi_{d}^2 q^2 + 1)^2\Big]\Big[\xi_{d}^2 q^2 + 1\Big]}\\
    &= \frac{\pi a v_0^2}{\tau}\delta(q+q')\frac{\zeta^2}{\zeta_q^2}\frac{\frac{\mu q^2}{\zeta_q}e^{\frac{-\mu q^2}{\zeta_q}|t-t'|} - \frac{1}{\tau}e^{-\frac{1}{\tau}|t-t'|}}{\left[\mu q^2/\zeta_q -1/\tau \right]\left[\mu q^2/\zeta_q +1/\tau \right]}.
\label{vv, qt space, steady state, resolved, pair-dissipation}
\end{aligned}
\end{equation}
where we have introduced
\begin{equation}
\zeta_q = \zeta + \eta q^2.
\end{equation}

\section{Analysis of the height equation}
\label{sec:height-analysis}

\subsection{Solving the height equation in $(q,t)$-space}

Using the Fourier transform conventions listed in Eq.\ \eqref{Fourier transform conventions}, the height equation, Eq.\ \eqref{height equation, pair-dissipation}, takes the following form in $(q,t)$-space
\begin{align}
    (\zeta+\eta q^2) \partial_t \tilde{h}(q,t) = - \left(\lambda q^2 + \kappa q^4 \right)\tilde{h}(q,t) + \zeta \tilde{v}^{f}(q,t).
\label{height equation, qt space}
\end{align}
which may be rewritten, using Eq.\ \eqref{eq:zetaq},
\begin{align}
     \partial_t \tilde{h} + \frac{\lambda q^2 + \kappa q^4}{\zeta_q}\tilde{h} = \frac{\zeta}{\zeta_q}\tilde{v}^{f}.
\label{height equation, de, qt space, pair dissipation}
\end{align}
Considering $e^{\left(\lambda q^2 + \kappa q^4 \right)t/\zeta_q}$ as integrating factor, we write
\begin{align}
     \partial_t \left(\tilde{h} e^{\left(\lambda q^2 + \kappa q^4 \right)t/\zeta_q}\right) = \frac{\zeta}{\zeta_q} \tilde{v}^{f}e^{\left(\lambda q^2 + \kappa q^4 \right)t/\zeta_q},
\label{height equation, IF, qt space}
\end{align}
%
%
%
and therefore by integration
\begin{align}
    \tilde{h}(q,t) = \frac{\zeta}{\zeta_q} \int_0^{t} \dd t' \, \tilde{v}^{f}(q,t')e^{-\left(\lambda q^2 + \kappa q^4 \right)(t-t')/\zeta_q},
\label{h, qt space, pair dissipation}
\end{align}
where we have used the initial condition $h(x,0)=0$ thus $\tilde{h}(q,0)=0$.

\paragraph*{Average border progression.}
Since $\la \hat{v}^{f}(q,t') \ra=0$, we have by linearity of the Fourier transform
\begin{subequations}
\begin{align}
    \la \tilde{h}(q,t)\ra &= 0,\\
    \Big< h(x,t)\Big> &= 0,
\end{align}
\end{subequations}
i.e.\ no average border progression.

\subsection{Transient Spatial Fourier spectrum}

Using equation Eq.\ \eqref{h, qt space, pair dissipation}, we get for the transient height spectrum
\begin{align}
    \la \tilde{h}(q,t) \tilde{h}(q',t)\ra = \frac{\zeta^2}{\zeta_q^2}\int_0^{t}\int_0^{t} \dd t_1 \, \dd t_2 \,\la\tilde{v}^{f}(q,t_1)\tilde{v}^{f}(q',t_2) \ra e^{- \left[\lambda (q^2 \left[t-t_1\right] + q'^2 \left[t-t_2\right]) + \kappa (q^4\left[t-t_1\right] + q'^4\left[t-t_2\right]) \right]/\zeta_q}.
    \label{hh, qt-space, implicit velocity correlations, pair-dissipation}
\end{align}
We assume that bulk velocity correlations have reached a steady state well before the release of the confinement.
This allows us to use the steady state result for $\la\tilde{v}^{f}(q,t_1)\tilde{v}^{f}(q',t_2) \ra$ Eq.\ \eqref{vv, qt space, steady state, resolved, pair-dissipation} and write
\begin{equation}
\begin{aligned}
 \la \tilde{h}(q,t) \tilde{h}(q',t)\ra &=  \frac{\pi a v_0^2}{\tau} \delta(q+q')\frac{\zeta^4}{\zeta_q^4}\int_0^{t}\int_0^{t} \dd t_1 \, \dd t_2\frac{\frac{\mu q^2}{\zeta_q}e^{\frac{-\mu q^2}{\zeta_q}|t_1-t_2|} - \frac{1}{\tau}e^{-\frac{1}{\tau}|t_1-t_2|}}{\left[\mu q^2/\zeta_q -1/\tau \right]\left[\mu q^2/\zeta_q +1/\tau \right]}e^{-\left[\lambda q^2 + \kappa q^4 \right]\left[2t-t_1-t_2\right]/\zeta_q},\\
 &=\frac{\pi a v_0^2 \delta(q+q')}{\tau \left[\mu q^2/\zeta_q -1/\tau \right]\left[\mu q^2/\zeta_q +1/\tau \right]}\frac{\zeta^4}{\zeta_q^4}H(q,t), 
\end{aligned}
\label{hh, qt-space}
\end{equation}
where we find
\begin{align}
    H(q,t) =& \left(1 - e^{-2\left[\lambda q^2 + \kappa q^4 \right]t/\zeta_q}\right) \frac{\lambda q^2 + \kappa q^4}{\zeta_q}\frac{(1/\tau)^2 - (\mu q^2/\zeta_q)^2}{ \left[\left[\lambda q^2 + \kappa q^4 \right]^2/\zeta_q^2 - \left(1/\tau \right)^2\right] \left[\left[\lambda q^2 + \kappa q^4 \right]^2/\zeta_q^2 - \left(\mu q^2/\zeta_q \right)^2 \right]} \nonumber\\[2ex]
    &+\left(1 +e^{-2\left[\lambda q^2 + \kappa q^4 \right]t/\zeta_q} -2e^{-\left[ \lambda q^2 + \kappa q^4\right]t/\zeta_q- \mu q^2t/\zeta_q} \right)\left[\frac{\mu q^2/\zeta_q}{\left[\lambda q^2 + \kappa q^4 \right]^2/\zeta_q^2 - \left(\mu q^2/\zeta_q \right)^2} \right] \\[2ex]
    &-\left( 1 +e^{-2\left[\lambda q^2 + \kappa q^4 \right]t/\zeta_q} -2 e^{-\left[ \lambda q^2 + \kappa q^4\right]t/\zeta_q - t/\tau}\right)\left[\frac{1/\tau}{\left[\lambda q^2 + \kappa q^4 \right]^2/\zeta_q^2 - \left(1/\tau \right)^2} \right]. \nonumber
\end{align}

\mbox{}\\
\paragraph*{Infinite time limit.}
For all $q^2 \neq 0$, we find
\begin{align}
    \lim_{t\rightarrow \infty}H(q,t) =& \frac{\lambda q^2 + \kappa q^4}{\zeta_q}\frac{(1/\tau)^2 - (\mu q^2/\zeta_q)^2}{ \left[\left[\lambda q^2 + \kappa q^4 \right]^2/\zeta_q^2 - \left(1/\tau \right)^2\right] \left[\left[\lambda q^2 + \kappa q^4 \right]^2/\zeta_q^2 - \left(\mu q^2/\zeta_q \right)^2 \right]} \nonumber \\[2ex]
    &+\left[\frac{\mu q^2/\zeta_q}{\left[\lambda q^2 + \kappa q^4 \right]^2/\zeta_q^2 - \left(\mu q^2/\zeta_q \right)^2} \right] \nonumber \\[2ex]
    &-\left[\frac{1/\tau}{\left[\lambda q^2 + \kappa q^4 \right]^2/\zeta_q^2 - \left(1/\tau \right)^2} \right]  \nonumber\\
    =& \frac{1}{(\lambda q^2 + \kappa q^4)/\zeta_q + 1/\tau} - \frac{1}{(\lambda q^2 + \kappa q^4)/\zeta_q + \mu q^2/\zeta_q}. 
\end{align}
This means that we can conclude 
\begin{align}
     \la \tilde{h}(q) \tilde{h}(q') \ra &\equiv\lim_{t\rightarrow \infty}\la \tilde{h}(q,t) \tilde{h}(q',t)\ra \nonumber \\[2ex] &= 
     \frac{\pi a v_0^2 \delta(q+q')}{\tau \left[\mu q^2/\zeta_q -1/\tau \right]\left[\mu q^2/\zeta_q +1/\tau \right]}\frac{\zeta^4}{\zeta_q^4} \left[\frac{1}{(\lambda q^2 + \kappa q^4)/\zeta_q + 1/\tau} - \frac{1}{(\lambda q^2 + \kappa q^4)/\zeta_q + \mu q^2/\zeta_q} \right],
     \label{hh, qq', infinite time limit, pairdis}
\end{align}
for any $q^2\neq 0$.

\subsection{Transient roughness}
We define the roughness $w(t)$ as the ensemble average of the standard deviation of the height at time $t$:

\begin{align}
    w^2(t) \equiv   \la \overline{ \left(h(x,t) - \overline{h(x,t)}    \right)^2} \ra 
     =  \la \overline{h^2(x,t)} \ra - \la  \left(\overline{h(x,t)}\right)^2 \ra,
\end{align}
where the overline $\overline{A}$ refers to the spatial average, and brackets $\langle A \rangle$ to the ensemble (noise) average.
By writing the spatial average as an integral, we see that we can exchange the order of averaging, so we use Eq.\ \eqref{hh, qt-space} to first find

\begin{align}
    \la h^2(x,t) \ra = & \frac{1}{(2 \pi)^2} \iint \dd q \, \dd q' \la \tilde{h}(q,t) \tilde{h}(q',t)\ra e^{-\ii qx} e^{-\ii q'x} \label{roughness formal}\\[2ex]
    = & \frac{1}{(2 \pi)^2 } \int \dd q \frac{\pi a v_0^2 }{\tau \left[\mu q^2/\zeta_q -1/\tau \right]\left[\mu q^2/\zeta_q +1/\tau \right]}\frac{\zeta^4}{\zeta_q^4}H(q,t) \nonumber
    \label{difficult integral}
\end{align}
where we have used the Dirac delta to evaluate the integral over $q'$. 
And then 

\begin{align}
      \la \overline{h^2(x,t)} \ra = \overline{ \la h^2(x,t) \ra} = \lim_{L \rightarrow \infty} \frac{1}{L} \int_{-L/2} ^{L/2} \dd x \la h^2(x,t) \ra =   \lim_{L \rightarrow \infty} \frac{1}{L} L \la h^2(t) \ra =  \la h^2(t) \ra.
\end{align}

By symmetry of the equations of motion of both the height and support chain and the fact that the noise is spatially uncorrelated, we expect the average spatial height of the boundary to be zero, hence

\begin{align}
    \la  \left(\overline{h(x,t)}\right)^2 \ra = 0.
\end{align}

So to calculate the roughness, we (numerically) calculate the integral 
\begin{align}
    \la h^2(t) \ra =\frac{1}{(2 \pi)^2 } \int \dd q \frac{\pi a v_0^2 }{\tau \left[\mu q^2/\zeta_q -1/\tau \right]\left[\mu q^2/\zeta_q +1/\tau \right]}\frac{\zeta^4}{\zeta_q^4}H(q,t)
\end{align}

with 

\begin{align}
    H(q,t)=& \left(1 - e^{-2\left[\lambda q^2 + \kappa q^4 \right]t/\zeta_q}\right) \frac{\lambda q^2 + \kappa q^4}{\zeta_q}\frac{(1/\tau)^2 - (\mu q^2/\zeta_q)^2}{ \left[\left[\lambda q^2 + \kappa q^4 \right]^2/\zeta_q^2 - \left(1/\tau \right)^2\right] \left[\left[\lambda q^2 + \kappa q^4 \right]^2/\zeta_q^2 - \left(\mu q^2/\zeta_q \right)^2 \right]} \nonumber \\[2ex]
    &+\left(1 +e^{-2\left[\lambda q^2 + \kappa q^4 \right]t/\zeta_q} -2e^{-\left[ \lambda q^2 + \kappa q^4\right]t/\zeta_q- \mu q^2t/\zeta_q} \right)\left[\frac{\mu q^2/\zeta_q}{\left[\lambda q^2 + \kappa q^4 \right]^2/\zeta_q^2 - \left(\mu q^2/\zeta_q \right)^2} \right] \\[2ex]
    &-\left( 1 +e^{-2\left[\lambda q^2 + \kappa q^4 \right]t/\zeta_q} -2 e^{-\left[ \lambda q^2 + \kappa q^4\right]t/\zeta_q - t/\tau}\right)\left[\frac{1/\tau}{\left[\lambda q^2 + \kappa q^4 \right]^2/\zeta_q^2 - \left(1/\tau \right)^2} \right]. \nonumber
\end{align}

\clearpage
\paragraph{Transient relative roughness correlation.}
We define the relative roughness as in \cite{rapinRoughnessDynamicsProliferating2021}
\begin{equation}
\begin{aligned}
    B_S(x, t) \equiv &  \la \overline{ \left(h(x+x',t) - h(x',t)\right)^2} \ra\\[2ex]
    =& 2 \la h^2(t) \ra - 2 \la\overline{ h(x+x', t)h(x',t)} \ra,
\end{aligned}
\end{equation}
where we numerically calculate
\begin{align}
    \la\overline{ h(x, t)h(x',t)} \ra=\frac{1}{(2 \pi)^2} \int \dd q
  \frac{\pi a v_0^2 }{\tau \left[\mu/\zeta_q -1/\tau \right]\left[\mu q^2/\zeta_q +1/\tau \right]}\frac{\zeta^4}{\zeta_q^4}H(q,t) e^{-\ii q (x-x')}.
\end{align}

\paragraph{Transient temporal roughness correlation.}
We define a temporal roughness correlation of the following form,

\begin{align}
    B_T(\Delta t, t) \equiv  \frac{ \langle \overline{\Delta h(x,t+\Delta t) \Delta h(x,t)} \rangle}{ \sqrt{\langle \overline{\Delta h(x,t+\Delta t)^2} \rangle} \sqrt{ \langle \overline{\Delta h(x,t)^2} \rangle}},
\end{align}
with $\Delta h (x,t) \equiv h(x,t) - \overline{h(x,t)}$. Noting that the terms in the denominator are just the variances of the height at the respective times, we can also write

\begin{align}
    B_T(\Delta t, t) = \frac{ \langle \overline{\Delta h(x,t+\Delta t) \Delta h(x,t)} \rangle}{ \sqrt{w^2(t+\Delta t)} \sqrt{w^2(t)} }.
\end{align}

We, therefore, need

\begin{align}
    \langle \tilde{h}(q,t) \tilde{h}(q',t') \rangle=\frac{\zeta^2}{\zeta_q^2} \int_0^{t}\int_0^{t'} \dd t_1 \dd t_2 \la\tilde{v}^{f}(q,t_1)\tilde{v}^{f}(q',t_2) \ra e^{- \left[\lambda (q^2 \left[t-t_1\right] + q'^2 \left[t'-t_2\right]) + \kappa (q^4\left[t-t_1\right] + q'^4\left[t'-t_2\right]) \right]/\zeta_q}.
\end{align}

If we now introduce $t' = t + \Delta t$, we find eventually

\begin{align}
    &\langle \tilde{h}(q,t) \tilde{h}(q',t + \Delta t) \rangle= \nonumber\\
    &\frac{\zeta^2}{\zeta_q^2} \int_0^{t}\int_0^{t + \Delta t} \dd t_1 \dd t_2 \la\tilde{v}^{f}(q,t_1)\tilde{v}^{f}(q',t_2) \ra e^{- \left[\lambda (q^2 \left[t-t_1\right] + q'^2 \left[t-t_2\right]) + \kappa (q^4\left[t-t_1\right] + q'^4\left[t-t_2\right]) \right]/\zeta_q} 
    e^{-\left[\lambda q'^2 + \kappa q'^4\right]\Delta t/\zeta_q}.
\end{align}

Plugging in the velocity correlations, one obtains (using Mathematica), the following result
\begin{align}
    \langle h(q,t) h(q',t + \Delta t) \rangle= \frac{ \pi a v_0^2}{\tau} \frac{\zeta^4}{\zeta_q^4} \zeta_q^3 \tau^2  \delta(q+q')G(q,t),
\end{align}
where $G(q,t, \Delta t) \equiv G_1(q,t, \Delta t) +  G_2(q,t, \Delta t)+  G_3(q,t, \Delta t) +  G_4(q,t, \Delta t)$, and

\begin{align*}
    &G_1(q,t, \Delta t) \equiv f_1(q) \left(e^{-\left[\lambda q^2 + \kappa q^4 \right]t/\zeta_q - (t+\Delta t)/\tau} + e^{-\left[\lambda q^2 + \kappa q^4 \right](t+\Delta t)/\zeta_q - \Delta t/\tau} -e^{-\Delta t/\tau}\right), \\
    &G_2(q,t, \Delta t) \equiv f_2(q) \left( e^{-\left[\lambda q^2 + \kappa q^4 \right](t+\Delta t)/\zeta_q - \mu q^2 t/\zeta_q}  +e^{-\left[(\lambda + \mu) q^2 + \kappa q^4 \right]t/\zeta_q - \mu q^2\Delta t/\zeta_q}- e^{-\mu q^2 \Delta t/\zeta_q}
    \right), \\
    &  G_3(q,t, \Delta t) \equiv f_3(q) e^{-\left[\lambda q^2 + \kappa q^4 \right]\Delta t/\zeta_q}, \\
    & G_4(q,t, \Delta t) \equiv f_4(q) e^{-\left[\lambda q^2 + \kappa q^4 \right](2t+\Delta t)/\zeta_q},
\end{align*}
where the $\{f_i(q)\}$ are rational functions of $q$.
By taking the inverse Fourier transform of $ \langle h(q,t) h(q',t + \Delta t) \rangle$ we then obtain $\langle \overline{\Delta h(x,t+\Delta t) \Delta h(x,t)} \rangle$.

\clearpage

\subsection{Transient Slope-slope and tangent-tangent correlation}

Denote the slope of the height function by
\begin{align}
    u(x,t) \equiv \partial_x  h(x,t), \quad
\text{and} \quad
    \tilde{u}(q,t) = -\ii q \tilde{h}(q,t).
\end{align}
The slope-slope correlation in $(q,t)$-space is thus, using Eq.\ \eqref{hh, qt-space},
\begin{align}
    \boxed{ \la \tilde{u}(q,t) \tilde{u}(q',t) \ra = \la -q q' \tilde{h}(q,t) \tilde{h}(q',t) \ra = q^2
  \frac{\pi a v_0^2 \delta(q+q')}{\tau \left[\mu q^2/\zeta_q -1/\tau \right]\left[\mu q^2/\zeta_q +1/\tau \right]}\frac{\zeta^4}{\zeta_q^4}H(q,t). }
    \label{slope-slope, qt-space, pairdis}
\end{align}

The slope-slope correlation can be used to approximate the tangent-tangent correlation, for which we need the real space correlation:

\begin{align}
    \la u(x,t) u(x',t) \ra =&\frac{1}{(2 \pi)^2} \iint \dd q \dd q' \la \tilde{u}(q,t) \tilde{u}(q',t) \ra e^{-\ii qx} e^{-\ii q'x'}\nonumber \\[2ex]
    =&\frac{1}{(2 \pi)^2} \int \dd q q^2 
  \frac{\pi a v_0^2 }{\tau \left[\mu q^2/\zeta_q -1/\tau \right]\left[\mu q^2/\zeta_q +1/\tau \right]}\frac{\zeta^4}{\zeta_q^4}H(q,t) e^{-\ii q (x-x')},  
\end{align}
where we have used the Dirac delta to resolve the integral over $q'$.
Here we see a similar integral to the integral of Eq.\ $\eqref{difficult integral}$.

\paragraph{Infinite time limit.}
Interestingly, using Mathematica, we find that, also for $q=0$,
\begin{align}
    \la \tilde{u}(q) \tilde{u}(q') \ra \equiv & \lim_{t\rightarrow \infty}\la \tilde{u}(q,t) \tilde{u}(q',t) \ra \nonumber \\[2ex] 
    = &\frac{ \pi a \tau \zeta^4 v_0^2 \delta(q+q')}{\left[\zeta  + \eta q^2 \right] \left[\lambda + \mu +\kappa q^2 \right] \left[ \zeta + \eta q^2 + \tau \lambda q^2 + \tau \kappa q^4   \right] \left[\zeta + \eta q^2 + \tau \mu q^2 \right] },
    \label{eq:slope-slope, q,t, infinite time}
\end{align}
which means that we can use contour integration to calculate the real space, infinite time, slope-slope correlation:
    
\begin{align}
    \la u(x) u(x') \ra = &\frac{1}{(2 \pi)^2}\iint \dd q \dd q' \la \tilde{u}(q) \tilde{u}(q') \ra e^{-\ii qx}e^{-\ii q'x'} \nonumber \\[2ex]
    =&\frac{a v_0^2 \tau \zeta^4}{4 \pi } \int \dd q \frac{e^{-\ii q(x-x')}}{\left[\zeta  + \eta q^2 \right] \left[\lambda + \mu +\kappa q^2 \right] \left[ \zeta + \eta q^2 + \tau \lambda q^2 + \tau \kappa q^4   \right] \left[\zeta + \eta q^2 + \tau \mu q^2 \right] }.
    \label{uu, x it, with pairdis, integral}
\end{align}
We will treat the $x-x'>0$ situation and use a contour integral of a contour in the bottom half of the complex plane. To reveal the poles, we expand the the denominators term by term:

\begin{align*}
    \frac{1}{\zeta + \eta q^2} = \frac{1}{\eta}\frac{1}{\zeta/\eta + q^2} = \frac{1}{\eta}\frac{1}{\left[q+ \ii \sqrt{\zeta /\eta} \right] \left[q -  \ii \sqrt{\zeta /\eta}\right]},
\end{align*}

\begin{align*}
    \frac{1}{\lambda + \mu + \kappa q^2} = \frac{1}{\kappa} \frac{1}{\lambda/\kappa + \mu /\kappa + q^2}=\frac{1}{\kappa}\frac{1}{\left[q+ \ii \sqrt{\frac{\lambda + \mu}{\kappa}} \right]  \left[ q - \ii \sqrt{\frac{\lambda + \mu}{\kappa}}\right]},
\end{align*}

\begin{align*}
    \frac{1}{\zeta + \eta q^2 + \tau \lambda q^2 + \tau \kappa q^4} = & \frac{1}{\kappa \tau}\frac{1}{\zeta/(\kappa \tau ) + \eta q^2 / (\kappa \tau) + \lambda q^2/\kappa + q^4} \\[2ex]
    =& \frac{1}{\kappa \tau} \frac{1}{\left[q-\ii \sqrt{-\frac{\lambda + \eta/\tau}{2 \kappa} - \sqrt{\left( \frac{\lambda + \eta/\tau}{2 \kappa} \right)^2 - \frac{\zeta}{\kappa \tau}}} \right]\left[q+\ii \sqrt{-\frac{\lambda + \eta/\tau}{2 \kappa} - \sqrt{\left( \frac{\lambda + \eta/\tau}{2 \kappa} \right)^2 - \frac{\zeta}{\kappa \tau}}} \right]} \\[2ex]
    &\cdot \frac{1}{\left[q-\ii\sqrt{-\frac{\lambda + \eta/\tau}{2 \kappa} +\sqrt{\left( \frac{\lambda + \eta/\tau}{2 \kappa} \right)^2 - \frac{\zeta}{\kappa \tau}}} \right]\left[q+\ii \sqrt{-\frac{\lambda + \eta/\tau}{2 \kappa} + \sqrt{\left( \frac{\lambda + \eta/\tau}{2 \kappa} \right)^2 - \frac{\zeta}{\kappa \tau}}} \right]},
\end{align*}
and finally,
\begin{equation*}
    \frac{1}{\zeta + \eta q^2 + \tau \mu q^2 } = \frac{1}{\eta + \mu \tau} \frac{1}{\frac{\zeta}{\eta + \mu \tau}+q^2} =\frac{1}{\eta + \mu \tau} \frac{1}{\left[q + \ii \sqrt{\frac{\zeta}{\eta + \mu \tau}} \right] \left[q - \ii \sqrt{\frac{\zeta}{\eta + \mu \tau}} \right]}.
\end{equation*}
The poles all lie on the imaginary axis if $\left(\frac{\lambda + \eta/\tau}{2 \kappa} \right)^2-\frac{\zeta}{\kappa \tau}\geq 0$, which is the situation for experiments and matched AVM simulations.
In this case, we can write the poles as
\begin{align}
    q_1 =q_2^*=& \ii \sqrt{\frac{\zeta}{\eta +\mu \tau}} \nonumber\\[2ex]
    q_3 = q_4^*=& \ii \sqrt{\left|\frac{\lambda+ \eta/\tau}{2 \kappa} + \sqrt{\left(\frac{\lambda+ \eta/\tau}{2 \kappa}\right)^2-\frac{\zeta}{\kappa \tau}}\right|} \nonumber \\[2ex]
    q_5 = q_6^*=& \ii \sqrt{\left|\frac{\lambda+ \eta/\tau}{2 \kappa} - \sqrt{\left(\frac{\lambda+ \eta/\tau}{2 \kappa}\right)^2-\frac{\zeta}{\kappa \tau}}\right|}  \label{eq:pole1-10}\\[2ex]
    q_7 = q_8^*=& \ii \sqrt{\frac{\lambda + \mu}{\kappa}} \nonumber \\[2ex]
    q_9 = q_{10}^*=& \ii \sqrt{\frac{\zeta}{\eta}}. \nonumber
\end{align}
In comparison to a previously studied height equation without pair dissipation, we note that we now have two extra poles, $q_9$ and $q_{10}$, representing a new length scale from the pair dissipation and that the active poles (containing $\tau$) are slightly modified.

With these abbreviations, we can write Eq.\ \eqref{uu, x it, with pairdis, integral} as

\begin{align}
    \la u(x) u(x') \ra =\frac{a v_0^2 \tau \zeta^4}{4 \pi } \frac{1}{\eta \kappa^2 \tau \left(\eta + \mu \tau \right)} \int \dd q \frac{e^{-iq(x-x')}}{\left[q- q_1 \right]\left[q- q_2 \right]\left[q- q_3 \right]\left[q- q_4 \right]\left[q- q_5 \right]\left[q- q_6 \right]\left[q- q_7 \right]\left[q- q_8 \right]\left[q- q_9 \right]\left[q- q_{10} \right]}.
\end{align}

For $x>x'$, we should take the contour in the lower half of the complex plane, such that $\mathrm{Im}[q]<0$. We take the clockwise contour and sum of the residues (while absorbing the minus sign from the clockwise contour) of the poles that lie in the bottom half (the even-numbered poles). It is then useful to write the integrand as 

\begin{align}
    \frac{e^{-\ii q(x-x')}}{\left[q-q_2^* \right] \left[q-q_2 \right] \left[q-q_4^* \right] \left[ q-q_4 \right]\left[q-q_6^* \right] \left[ q-q_6 \right]\left[q-q_8^* \right] \left[ q-q_8 \right]\left[q-q_{10}^* \right] \left[ q-q_{10} \right]}.
\end{align}

\begin{align*}
    \mathrm{Res}=& -2 \pi \ii  \frac{e^{-\ii q_2 (x-x')}}{\left[2q_2 \right] \left[q_2^2 - q_4^2 \right]\left[q_2^2 - q_6^2 \right]\left[q_2^2 - q_8^2 \right]\left[q_2^2 - q_{10}^2 \right]} \\[2ex]
    &-2\pi \ii \frac{e^{-\ii q_4 (x-x')}}{\left[q_4^2 - q_2^2 \right] \left[2 q_4 \right]\left[q_4^2 - q_6^2 \right]\left[q_4^2 - q_8^2 \right]\left[q_4^2 - q_{10}^2 \right] } \\[2ex]
    &-2 \pi \ii \frac{e^{-\ii q_6 (x-x')}}{\left[q_6^2 - q_2^2 \right]\left[q_6^2 - q_4^2 \right]\left[2 q_6\right]\left[q_6^2 - q_8^2 \right]\left[q_6^2 - q_{10}^2 \right]} \\[2ex]
    &-2 \pi \ii \frac{e^{-\ii q_8 (x-x')}}{\left[q_8^2 - q_2^2 \right]\left[q_8^2 - q_4^2 \right]\left[q_8^2 - q_6^2 \right]\left[2 q_8\right]\left[q_8^2 - q_{10}^2 \right]} \\[2ex]
    &-2 \pi \ii \frac{e^{-\ii q_{10} (x-x')}}{\left[q_{10}^2 - q_2^2 \right]\left[q_{10}^2 - q_4^2 \right]\left[q_{10}^2 - q_6^2 \right]\left[q_{10}^2 - q_{8}^2 \right]\left[2 q_{10}\right]}.
\end{align*}

In terms of the magnitudes of the poles we get, after defining $R(x-x') \equiv \mathrm{Res}/\pi$,

\begin{align*}
   R(x-x')= & \frac{e^{-|q_2| (x-x')}}{|q_2|  \left[|q_2|^2 - |q_4|^2 \right]\left[|q_2|^2 - |q_6|^2 \right]\left[|q_2|^2 - |q_8|^2 \right]\left[|q_2|^2 - |q_{10}|^2 \right]} \\[2ex]
    &+\frac{e^{-|q_4| (x-x')}}{\left[|q_4|^2 - |q_2|^2 \right] |q_4 |\left[|q_4|^2 - |q_6|^2 \right]\left[|q_4|^2 - |q_8|^2 \right]\left[|q_4|^2 - |q_{10}|^2 \right] } \\[2ex]
    &+\frac{e^{-|q_6| (x-x')}}{\left[|q_6|^2 - |q_2|^2 \right]\left[|q_6|^2 - |q_4|^2 \right]|q_6|\left[|q_6|^2 - |q_8|^2 \right]\left[|q_6|^2 - |q_{10}|^2 \right]} \\[2ex]
    &+ \frac{e^{-|q_8| (x-x')}}{\left[|q_8|^2 - |q_2|^2 \right]\left[|q_8|^2 - |q_4|^2 \right]\left[|q_8|^2 - |q_6|^2 \right]|q_8|\left[|q_8|^2 - |q_{10}|^2 \right]} \\[2ex]
    &+ \frac{e^{-|q_{10}| (x-x')}}{\left[|q_{10}|^2 - |q_2|^2 \right]\left[|q_{10}|^2 - |q_4|^2 \right]\left[|q_{10}|^2 - |q_6|^2 \right]\left[|q_{10}|^2 - |q_{8}|^2 \right]|q_{10}|},
\end{align*}
which means we find the infinite-time slope-slope correlation in real space for $x>x'$,
\begin{align}
    \la u(x) u(x') \ra =\frac{a v_0^2 \tau \zeta^4}{4} \frac{1}{\eta \kappa^2 \tau \left(\eta + \mu \tau \right)}R(x-x').
\end{align}
Note that the slope-slope correlation decays with 5 different decay lengths.
\clearpage

\paragraph{From slope-slope correlation to tangent-tangent correlation.}

The tangent-tangent correlation function is defined as

\begin{align}
   C(s,t)\equiv \la \overline{ \buv{t}(s+s',t) \cdot \buv{t}(s',t)}\ra,
\end{align}
where the spatial average is over $s'$.
If we assume the absence of overhangs (as we did for the derivation of the height equation), we can write $y(s) = y(s(x))$ and write the tangent vector in components:

\begin{align}
    \buv{t}(s,t) = &\partial_s x(s) \buv{x} + \partial_s y(s) \buv{y} \nonumber \\[2ex]
     = & \frac{\partial x}{\partial s} \left[ \buv{x} + \frac{\partial y}{\partial x}  \buv{y}\right].
\end{align}

We can then write the tangent-tangent product as

\begin{align}
    \buv{t}(s+s',t) \cdot \buv{t}(s',t) = \frac{\partial x}{\partial s} \bigg\rvert_{s+s'} \frac{\partial x}{\partial s} \bigg\rvert_{s'} \left[1 + \frac{\partial y}{\partial x} \bigg\rvert_{s+s'}\frac{\partial y}{\partial x} \bigg\rvert_{s'} \right].
\end{align}

Since $ds = \sqrt{dx^2 + dy^2}$, we have $\frac{\partial x}{ \partial s} = \left[1+ \left(\frac{\partial y}{\partial x}\right)^2 \right]^{-1/2} = 1 - \frac{1}{2} \left(\frac{\partial y}{\partial x}\right)^2 + \mathrm{h.o.t.}$  and so

\begin{align}
    \frac{\partial x}{\partial s} \bigg\rvert_{s+s'} \frac{\partial x}{\partial s} \bigg\rvert_{s'} \approx 1 - \frac{1}{2} \left[\frac{\partial y}{\partial x}\right]_{s+s'}^2 - \frac{1}{2} \left[\frac{\partial y}{\partial x}\right]_{s'}^2.
\end{align}

Therefore, 

\begin{align}
    \overline{ \buv{t}(s+s',t) \cdot \buv{t}(s',t)}  \approx & \overline{\left(1 - \frac{1}{2} \left[\frac{\partial y}{\partial x}\right]_{s+s'}^2 - \frac{1}{2} \left[\frac{\partial y}{\partial x}\right]_{s'}^2 \right) \left( 1 +\frac{\partial y}{\partial x} \bigg\rvert_{s+s'}\frac{\partial y}{\partial x} \bigg\rvert_{s'} \right)} \nonumber\\[2ex]
    \approx &  \overline{1 - \frac{1}{2} \left[\frac{\partial y}{\partial x}\right]_{s+s'}^2 - \frac{1}{2} \left[\frac{\partial y}{\partial x}\right]_{s'}^2 + \frac{\partial y}{\partial x} \bigg\rvert_{s+s'}\frac{\partial y}{\partial x} \bigg\rvert_{s'} } \nonumber \\[2ex]
     &= 1 - \overline{\frac{\partial y}{\partial x} \bigg\rvert_{s'}\frac{\partial y}{\partial x} \bigg\rvert_{s'}} + \overline{\frac{\partial y}{\partial x} \bigg\rvert_{s+s'}\frac{\partial y}{\partial x} \bigg\rvert_{s'} }.
\end{align}

And thus, since the derived slope-slope correlation is already the spatial-average (it only depends on the difference $x$ (or $s$).

\begin{align}
    C(s,t)= \la \overline{ \buv{t}(s+s',t) \cdot \buv{t}(s',t)}\ra \approx 1 - \la u(0)u(0) \ra + \la u(x(s))u(x(0)) \ra.
\end{align}
To the same order of the previous approximation, we approximate $s \approx \left(1 + \frac{1}{2} \la u(0)u(0)\ra\right)x$, so that we obtain

\begin{align}
    C(s,t)\approx 1 - \la u(0)u(0) \ra + \la u\left(\frac{s}{1 + \frac{1}{2} \la u(0)u(0)\ra}\right)u(0) \ra.
\end{align}

\clearpage

\part{Experiments}
\setcounter{section}{0}

\section{Curation}
Unfortunately, not all 24 image sequences were suitable for further analysis of the boundary and interior, mainly because some of the wells did not reach confluence before the gasket was removed or because the peeling off of the gasket also removed some cells near the border.
From inspection of the initial image after gasket removal, we curated the experiments: from the 24 original image sequences, we were able to use 17 sequences for further analysis. 

\begin{figure}[h]
    \centering
    \includegraphics[width=1\linewidth]{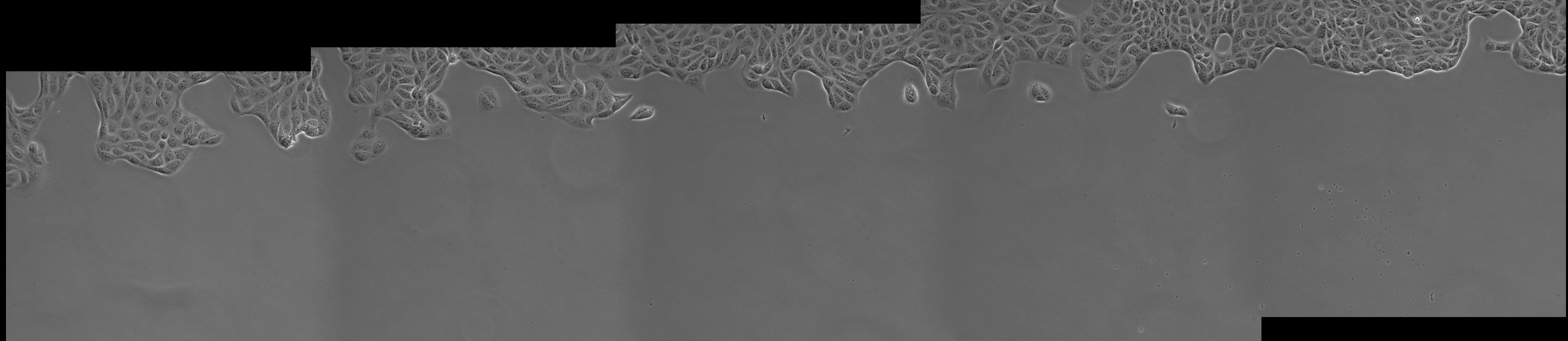}
    \caption{Example of an initial image  of a an image sequence we could not process further: image set 05, which shows clear sign of non-confluency: an initially non-straight edge and holes (see e.g. top left) in the tissue.}
    \label{fig:not_suitable_series}
\end{figure}

\begin{figure}[h]
    \centering
    \includegraphics[width=1\linewidth]{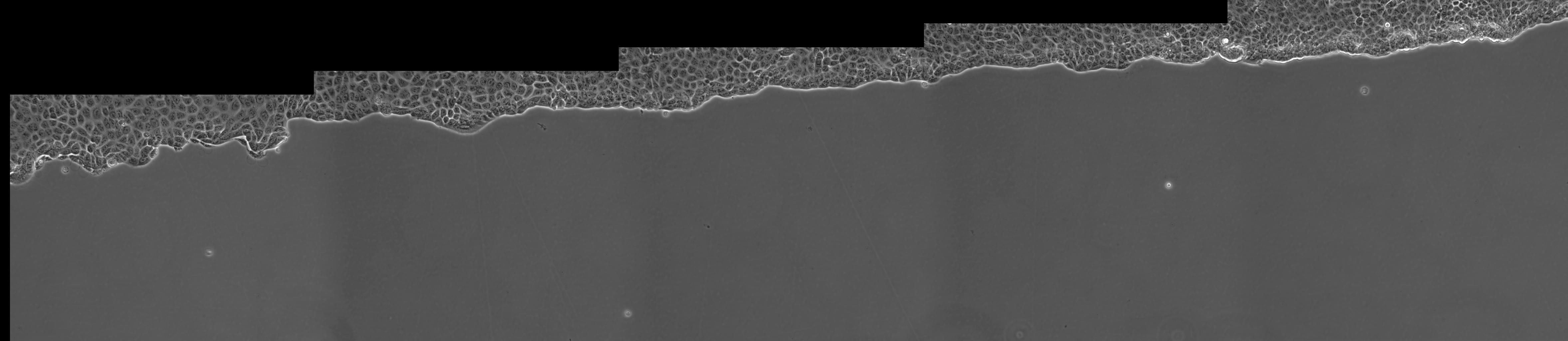}
    \caption{Example of an initial image  of an image sequence we processed: image set 01, which appears to be confluent, undamaged and straight.}
    \label{fig:suitable_series}
\end{figure}

For the remaining 17 sequences, we determined the final frame in which the boundary line could reasonably be extracted. This was determined either by the moving front having reached the other side of the field of view (even with the automatic following) or by some debris that started to touch the boundary as the front moved towards it, rendering any clean boundary detection in later frames ambiguous. An overview of the curation of the experiments is listed in Tab.\ \ref{table:curation}. 

\begin{table}[h!]
\begin{center}
\begin{tabular}{ c| c| c| c| c|  }
Experiment &	Acceptable initial frame &	Acceptable final frame & Orientation (L/R) & Include experiment (0/1)? \\
01 &	0000	&0362	&L	&1 \\
02 &	0000	&0188	&R	&1 \\
03 &	0000	&0190	&L	&1 \\
04 &	0000	&0000	&R	&0 \\
05 &	0000	&0000	&L	&0 \\
06 &	0000	&0000	&R	&0 \\
07 &	0000	&0000	&L	&0 \\
08 &	0000	&0473	&R	&1 \\
09 &	0000	&0092	&L	&1 \\
10 &	0000	&0054	&R	&1 \\
11 &    0000	&0000	&L	&0 \\
12 &    0000	&0287	&R	&1 \\
13 &    0000	&0150	&R	&1 \\
14 &    0000	&0114	&L	&1 \\
15 &	0000	&0087	&R	&1 \\
16 &    0000	&0283	&L	&1 \\
17 &	0000	&0000	&R	&0 \\
18 &	0000	&0287	&L	&1 \\
19 &	0000	&0236	&R	&1 \\
20 &	0000	&0418	&L	&1 \\
21 &	0000	&0426	&R	&1 \\
22 &	0000	&0080	&L	&1 \\
23 &	0000	&0000	&R	&0 \\
24 &	0000	&0350	&L	&1
\end{tabular}
\end{center}
\caption{Curation table showing which experiments are included and what the end frame is.}
\label{table:curation}
\end{table}

\clearpage
\section{Image processing}\label{image_processing}
Having curated the experimental dataset, the next step was to (pre)process the images so that the boundary line could be extracted and the PIV could be performed. Since the imaging setup was such that it followed the boundary over time, the images were essentially taken in a comoving reference frame with respect to the moving cell front. As we are interested in the absolute movement of the cell front, we needed to shift the images to the lab frame. In addition, we occasionally noticed a small horizontal drift of the field of view in the image sequences which we attribute to a small microscope drift. We corrected for this and the comoving frame by tracking a reference point (one of the corners in the field of view) in the images and using the location of this point to translate every image to an absolute reference frame, see figures \ref{fig:stabilizing_0000} and \ref{fig:stabilizing_0001}.

\begin{figure}[h]
    \centering
    \includegraphics[width=1\linewidth]{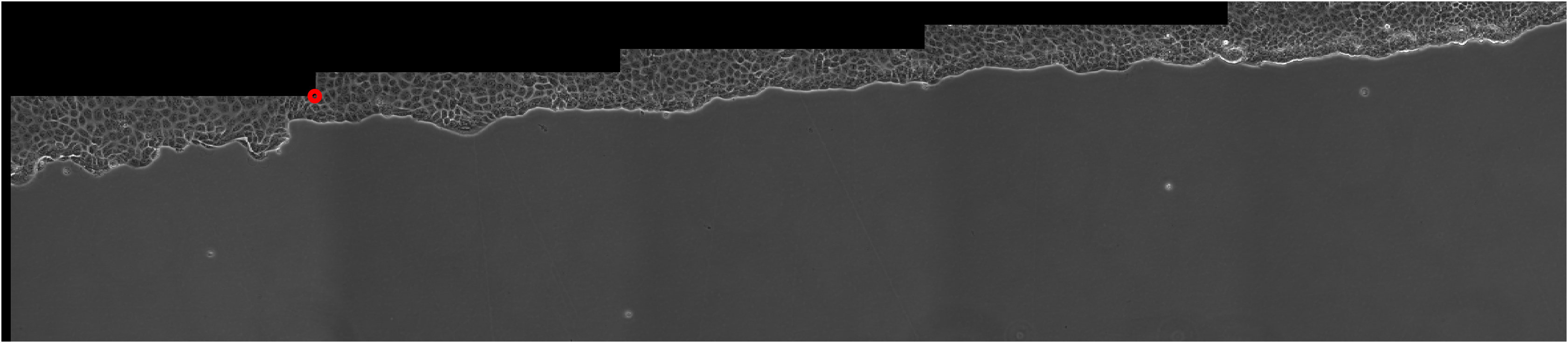}
    \caption{Frame 0000 of experiment 01, with the reference point indicated as the red circle}
    \label{fig:stabilizing_0000}
\end{figure}

\begin{figure}[h]
    \centering
    \includegraphics[width=1\linewidth]{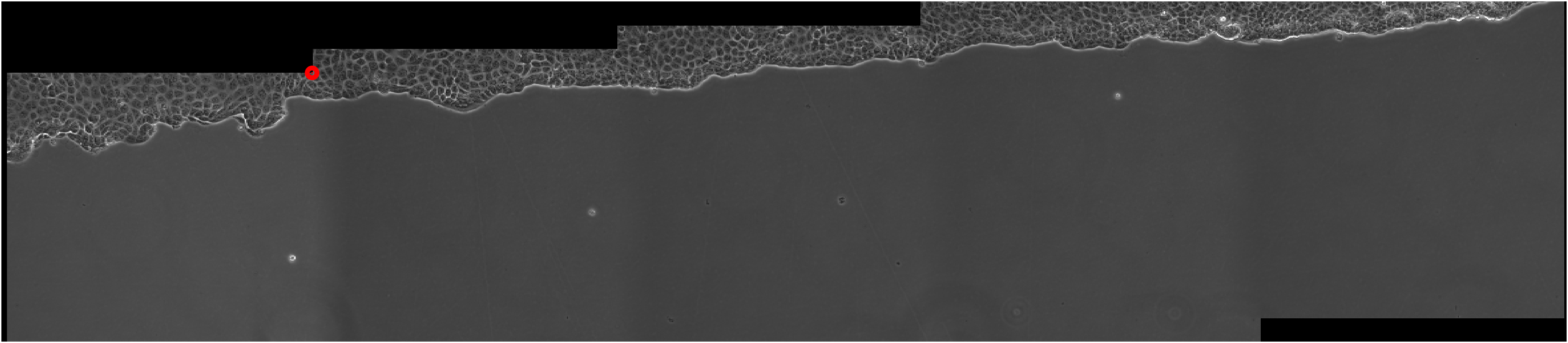}
    \caption{Frame 0001 of experiment 01, with the same reference point as in the previous frame. Note how the reference points has shifted upwards and slightly sideways.}
    \label{fig:stabilizing_0001}
\end{figure}

Using a number of subsequent morphological operations offered by the MATLAB image processing toolbox, we extracted the boundary pixels (see figure \ref{fig:boundary_extraction_0100}) and the tissue pixels (see figure \ref{fig:pas_0100}).
\begin{figure}[h]
    \centering
    \includegraphics[width=1\linewidth]{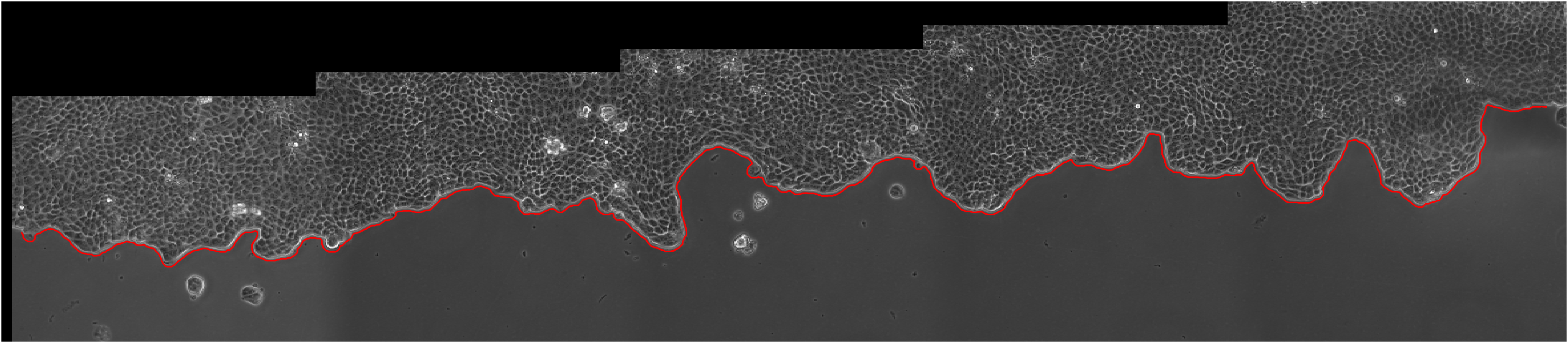}
    \caption{Frame 0100 of experiment 14. This image shows the extraction of the boundary, which is robust to the debris in the neighborhood.}
    \label{fig:boundary_extraction_0100}
\end{figure}

\begin{figure}[h]
    \centering
    \includegraphics[width=1\linewidth]{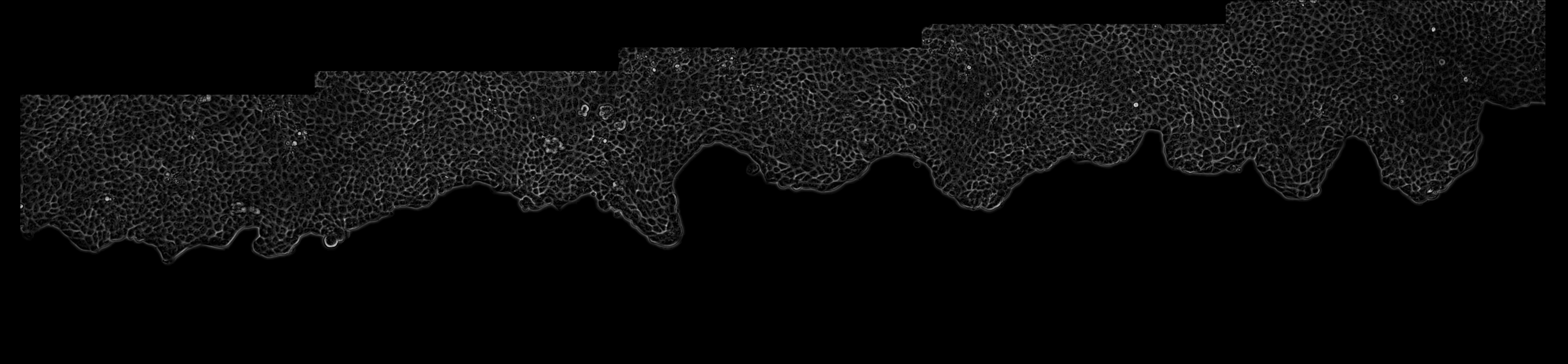}
    \caption{Frame 0100 of experiment 14. This is the resulting image after removing the uneven background,  setting the non-tissue pixels to zero and shifting to the lab frame.}
    \label{fig:pas_0100}
\end{figure}
To obtain a good quality PIV analysis, we (i) removed the uneven background (effect from the stitching), (ii) set all non-tissue pixels to zero, and (iii) shifted the image to the respective place in the absolute lab frame, see figure \ref{fig:pas_0100}.
\clearpage

\section{Data analysis}

\subsection{Zoning}\label{subsec:zoning}
Earlier experiments, e.g. \cite{petitjeanVelocityFieldsCollectively2010,reffayInterplayRhoAMechanical2014,vishwakarmaMechanicalInteractionsFollowers2018} revealed gradients in the velocity and force fields as a function of the distance from the euclidean distance from a fingertip. To analyze this in our experiments we divided the tissue in the images into dynamic zones, see figure \ref{fig:zoning}. The zones move in synchrony with the fingers, so that we could obtain velocity statistics at different locations behind the fingers. The finger region is determined from the maximum and minimum of the boundary line at any time. As the minimum position of the boundary is subject to (slow) fluctuations, we define a buffer zone to cleanly distinguish a finger zone and a region further behind the region between the minimum of the boundary line and 100 pixels below the mean of the boundary line ($\SI{1}{\micro\meter}\mathrel{\widehat{=}} 1.5385$ pixels). The velocity statistics of the region further away from the boundary than the buffer zone is shown in \ref{fig:deeper_in_bulk_statisticsl}.
\begin{figure}[h!]
    \centering
    \includegraphics[width=1\linewidth]{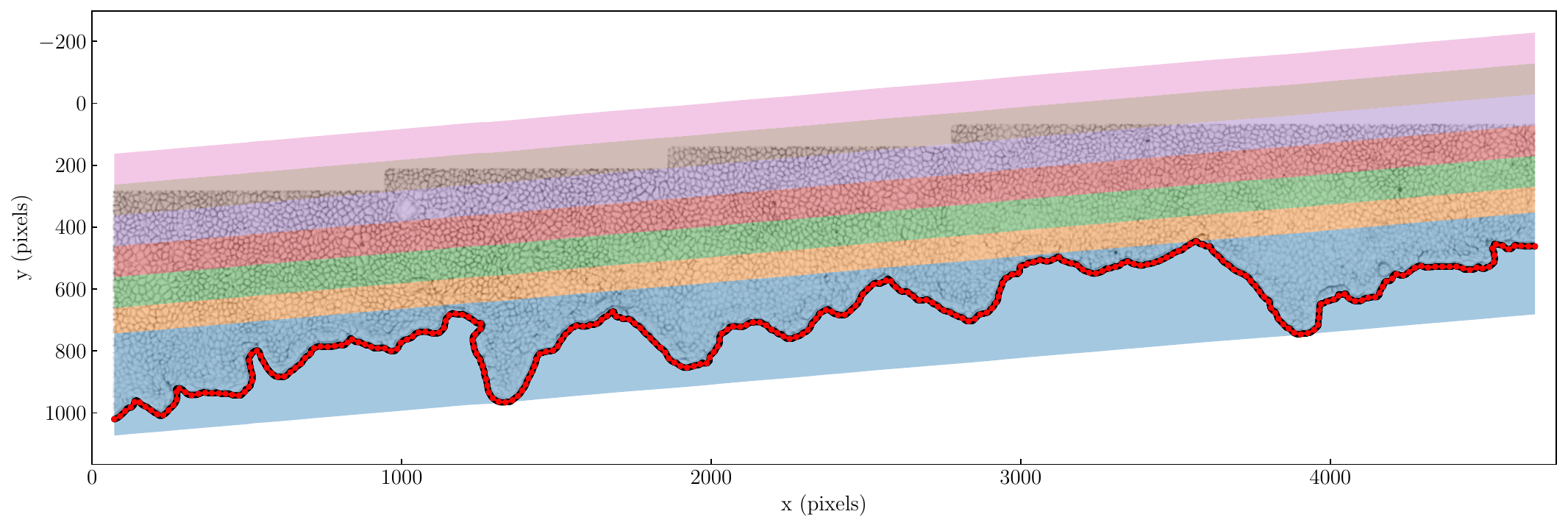}
    \caption{Frame 300 from experiment 01, inverted colors. The black points are the extracted boundary pixels from the image processing step. The red points are the spline smoothed points on this boundary that coarse grain the pixel nature of the boundary. The shaded regions are the different zones: blue is the finger region, orange is the buffer zone (difference between minimum of boundary line and a fixed distance (100 pixels) from the mean of the boundary line), the subsequent colors (green to pink) represent the different bulk zones of fixed width (also 100 pixels). The green to pink zones are also collectively analyzed and refered to as the interior. We show here the boundary without the rotational tilt correction to plot it on top of the microscope image of the tissue. }
    \label{fig:zoning}
\end{figure}

\subsection{PIV}\label{subsec:PIV}
We followed the work by Petitjean et al\cite{petitjeanVelocityFieldsCollectively2010} in determining the window size for the PIV: the maximal local displacement should be smaller than half of the size of the smallest window.
Like  Petitjean et al\cite{petitjeanVelocityFieldsCollectively2010}, we took  $30 \mathrm{\mu m}$/h as an overestimation of the speed of the cells. Since the time between frames is 5 minutes and $\SI{1}{\micro\meter}\mathrel{\widehat{=}} 1.5385$ pixels, the maximal displacement is 3.8 pixels between frames. As this is very close to 4 pixels, we decided to use 16 pixels and not 8 as the smallest window size because this still gives sub cellular size resolution (average cell area is around $\SI{170}{\micro\meter^2}$). The PIV analysis is carried out with the use of PIVlab\cite{thielickeParticleImageVelocimetry2021} for MATLAB with successively smaller window sizes: 64 to 32 to 16 pixels. We show in figure \ref{fig:zoning_PIV} an excerpt of the PIV field.
\begin{figure}[h]
    \centering
    \includegraphics[width=0.3\linewidth]{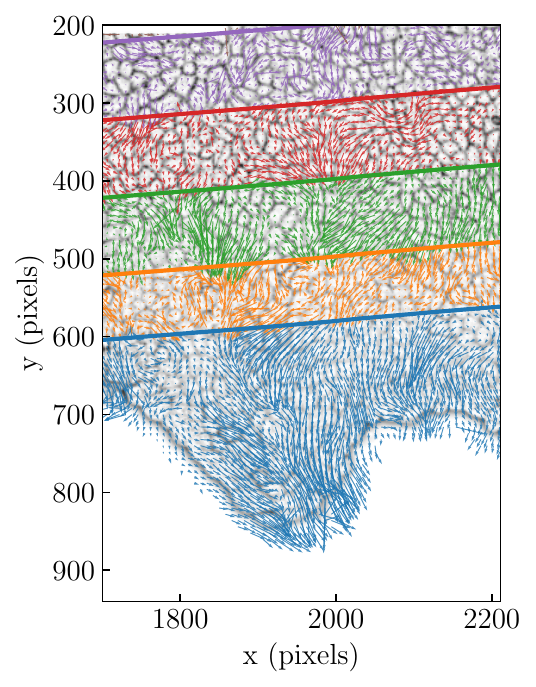}
    \caption{Frame 300 from experiment 01, inverted colors. The vectors are scaled PIV vectors. See figure \ref{fig:zoning} for the color coding.}
    \label{fig:zoning_PIV}
\end{figure}

\subsection{Rotational tilt correction}
Most experiments started with a slanted boundary, see for example the initial frame of experiment 01 in figure \ref{fig:stabilizing_0000}. We corrected for this by, for each experiment separately, fitting a linear line through the boundary of the first frame and using the angle of this line to rotate the boundary coordinates in all frames. We also did this for the position and velocity vectors of the PIV vectors. We validated the rotational tilt correction by measuring the average speed in the horizontal direction (along the long tissue axis) and the vertical direction before and after the rotational tilt correction. The rotational tilt correction led to an increase in the vertical speed in exchange for a smaller horizontal speed. This is consistent with the fact that initially the movement of the border cells is mostly oriented towards the free space \cite{poujadeCollectiveMigrationEpithelial2007}.

\subsection{Spline smoothing and regular grid binning}\label{spline_smoothing_grid_binning}
We also had to account for the pixel nature of the extracted boundary line. For example, if we were to use the boundary pixels directly as coordinates, the tangent vectors would only come in three variants: vertical, horizontal or at  \SI{45}{\degree} angles. To account for that, we coarse grained the boundary with a smoothing spline to roughly the scale of a single cell, see Figs.\ \ref{fig:spline_smoothing} and \ref{fig:spline_smoothing_zoomed}.

\begin{figure}[h]
    \centering
    \includegraphics[width=1\linewidth]{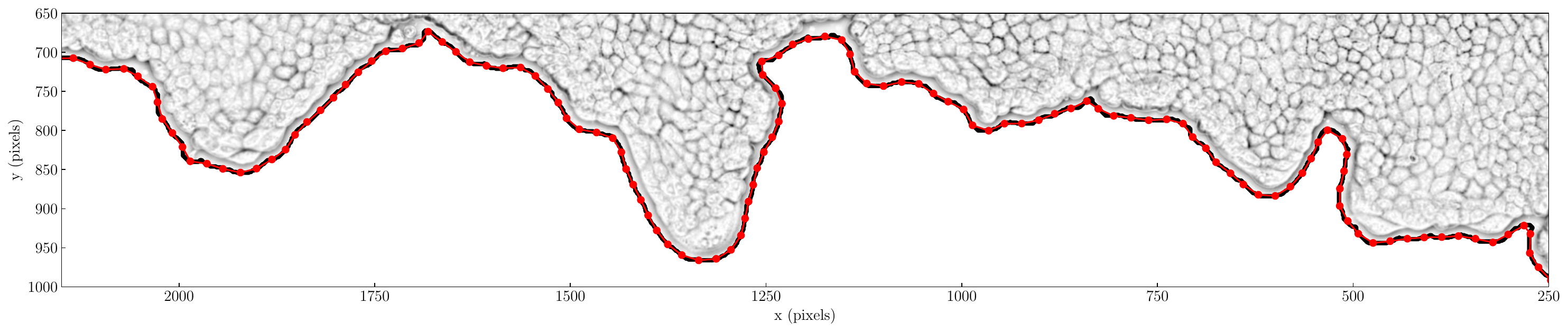}
    \caption{Frame 300 from experiment 01, inverted colors. The black points are the extracted boundary pixels from the image processing step. The red points are the spline smoothed points on this boundary that coarse grain the pixel nature of the boundary.}
    \label{fig:spline_smoothing}
\end{figure}

\begin{figure}[h]
    \centering
    \includegraphics[width=0.5\linewidth]{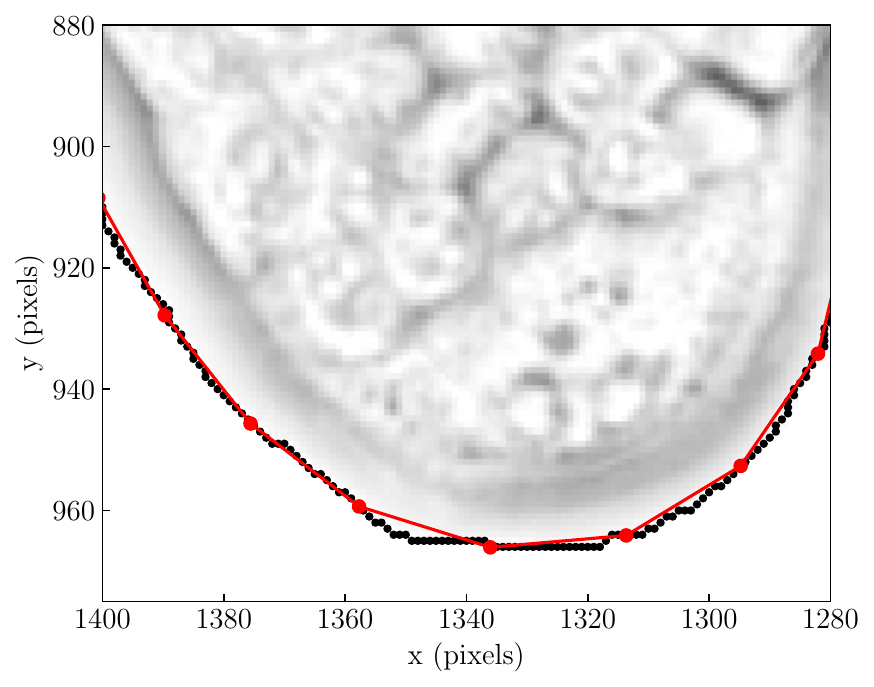}
    \caption{Frame 300 from experiment 01, inverted colors. The black points are the extracted boundary pixels from the image processing step. The red points are the spline smoothed points on this boundary that coarse grain the pixel nature of the boundary.}
    \label{fig:spline_smoothing_zoomed}
\end{figure}

To calculate the one-dimensional spatial Fourier transform of the boundary, we need unique and regularly spaced $x$-coordinates for each $y$-coordinate.
We achieved this by binning the $x$-axis and taking the maximal $y$-value in this bin as the $y$ value at the $x$ bin center. This also takes care of overhangs in the sense that it is extracting the convex hull of the boundary. See Fig.\ \ref{fig:regular_grid} for an example.
\begin{figure}[h]
    \centering
    \includegraphics[width=1\linewidth]{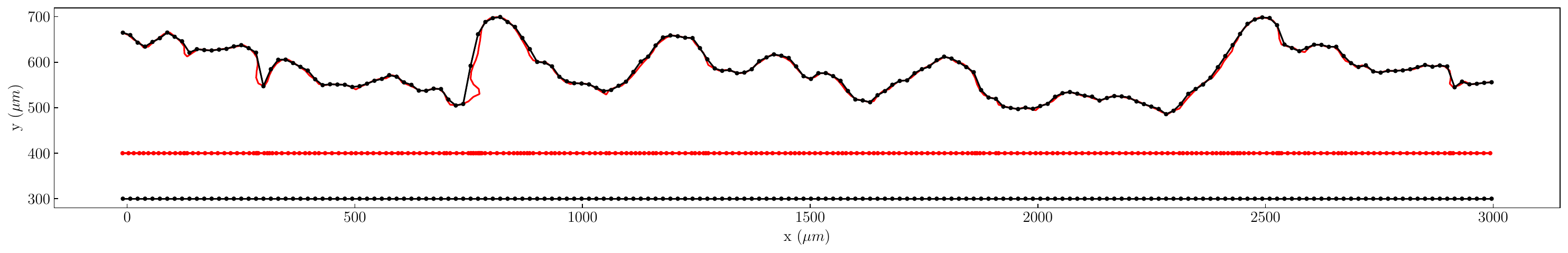}
    \caption{Boundary from frame 300 of experiment 01. The curvy red line is the resulting boundary line after the rotational tilt correction and the spline smoothing. The horizontal red line with markers shows the $x$-coordinates of the red boundary line. Clearly, the interval between these points is uneven. The black points from the regular grid interpolation of the boundary line with evenly spaced $x$-coordinates is visible as the curvy black line. The corresponding evenly spaced $x$-coordinates are shown in the horizontal black line. }
    \label{fig:regular_grid}
\end{figure}
\clearpage
\subsection{Segmentation}\label{segmentation}
As a starting point for the matching with AVM simulations, it is useful to know the distribution of area, perimeter and shape index. To extract these quantities from the experimental images, we segmented an excerpt from frame 350 of experiment 07 Note: this is an experiment that is excluded from the boundary dynamics analysis because it was not confluent at the start (frame 0).
The segmentation, shown in Fig.\ \ref{fig:exp_segmentation} is performed with MATLAB using the watershed algorithm after some filtering steps. 

\begin{figure}[h]
    \centering
    \includegraphics[width=0.4\linewidth]{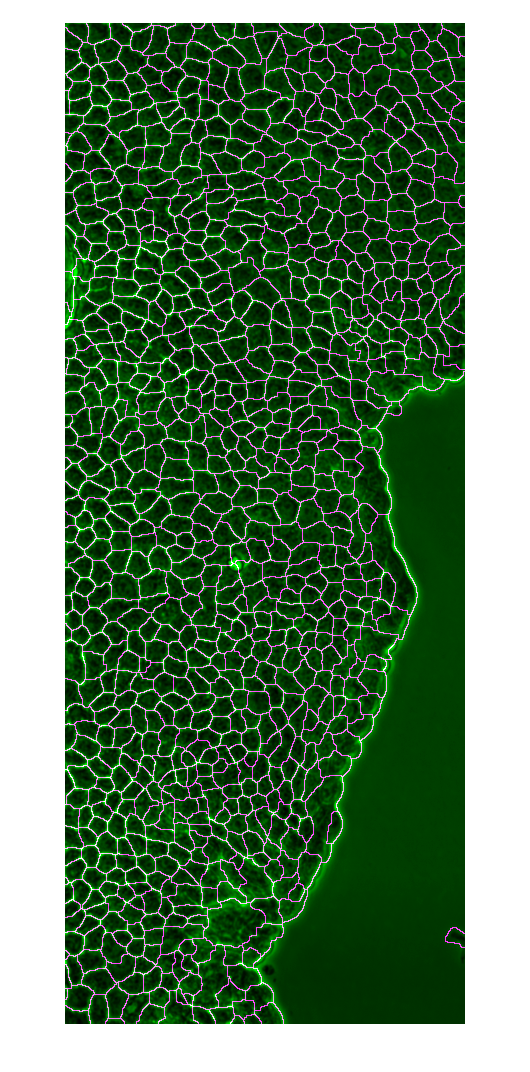}
    \caption{Segmentation of excerpt of frame 350 of experiment 07.}
    \label{fig:exp_segmentation}
\end{figure}

\end{justify}
\clearpage

\end{document}